\documentclass[fleqn,usenatbib]{mnras}

\usepackage[T1]{fontenc}
\usepackage{ae,aecompl}
\usepackage[thinc]{esdiff}
\usepackage{chemformula}
\usepackage[version=4]{mhchem}
\usepackage[utf8]{inputenc}
\usepackage{CJKutf8}

\usepackage{multirow}
\usepackage{hyperref}

\usepackage[normalem]{ulem}

\usepackage{graphicx}	% Including figure files
\usepackage{amsmath}	% Advanced maths commands
\usepackage{amssymb}	% Extra maths symbols

\usepackage{isotope}
\usepackage{siunitx}
\usepackage{wrapfig,framed}
\usepackage{outlines}

\newcommand{\approptoinn}[2]{\mathrel{\vcenter{
  \offinterlineskip\halign{\hfil$##$\cr
    #1\propto\cr\noalign{\kern2pt}#1\sim\cr\noalign{\kern-2pt}}}}}

\newcommand{\appropto}{\mathpalette\approptoinn\relax}

\title[ ]{Chemical evolution of an evaporating lava pool}

\author[]{Alfred Curry$^{1}$, Subhanjoy Mohanty$^{1}$, James E. Owen$^{1}$
\\
$^{1}$Astrophysics Group, Department of Physics, Imperial College London, Prince Consort Rd, London SW7 2AZ, UK}

\date{Accepted XXX. Received YYY; in original form ZZZ}

\pubyear{2022}

\begin{document}
\label{firstpage}
\pagerange{\pageref{firstpage}--\pageref{lastpage}}
\maketitle

\begin{abstract}
Many known rocky exoplanets are so highly irradiated that their dayside surfaces are molten, and `silicate atmospheres', composed of rock-forming elements, are generated above these lava pools. The compositions of these `lava planet' atmospheres are of great interest because they must be linked to the composition of the underlying rocky interiors. It may be possible to investigate these atmospheres, either by detecting them directly via emission spectroscopy or by observing the dust tails which trail the low mass `catastrophically evaporating planets'. In this work, we develop a simple chemical model of the lava pool--atmosphere system under mass loss, to study its evolution. Mass loss can occur both into space and from the day to the nightside. We show that the system reaches a steady state, where the material in the escaping atmosphere has the same composition as that melted into the lava pool from the mantle. We show that the catastrophically evaporating planets are likely to be in this evolved state. This means that the composition of their dust tails is likely to be a direct trace of the composition of the mantle material that is melted into the lava pool. We further show that, due to the strength of day-to-nightside atmospheric transport, this evolved state may even apply to relatively high-mass planets ($\gtrsim 1 M_\oplus$). Moreover, the low pressure of evolved atmospheres implies that non-detections may not be due to the total lack of an atmosphere. Both conclusions are important for the interpretation of future observations.
\end{abstract}

\section{Introduction}\label{sec:intro}

Since the discovery of the first super-Earths \citep{leger2009} there has been interest in the nature of their surfaces and atmospheres. Many super-Earths are believed to have surface temperatures well above the melting point of rocky material \citep[see, for instance,][Fig. 1]{zilinskas22} and thus host surface lava pools. Due to their extreme irradiation, these hot planets are likely to have lost the volatile elements that make up solar system atmospheres (H,C,N etc.) and thus should host `silicate atmospheres' made from conventionally rock-forming elements \citep{Schaefer_fegley2009}. 

These `lava planets' are of particular interest in the wider exoplanet field because the elements forming their atmospheres are usually locked in the solid parts of planets and so are very challenging to observe. Bulk compositions of rocky bodies can be observed in white dwarf atmospheres \citep[e.g.,][]{Zuckerman2010}, and so using lava atmospheres offers an interesting synergy with this information as the material observed in an atmosphere must have gone through some processing close to the surface. Furthermore, the precise origin of the pollutant material in white dwarfs is not yet clear, so those systems may probe a different type of planet \citep{Amyreview,buchan2021,Veras-delivery2024} or the composition may be altered by the accretion process \citep{Brouwers2023}. It may be possible to probe surface compositions through emission/reflection spectra \citep{Hu2012} but this is challenging and information from lava atmospheres will be complementary. Rocky planet interiors are greatly influenced by early planet evolution, while themselves influencing the composition and evolution of secondary atmospheres \citep[e.g.,][]{PP7-licht}. Therefore, the understanding of rocky planet interiors is essential for understanding planet diversity, including how common Earth-like conditions might be. 

The first study of `silicate atmospheres' was conducted by \citet{Schaefer_fegley2009}. They modelled the atmosphere generated by high-temperature lava using a code \texttt{MAGMA} \citep{fegley1987}, which was developed to study the high-temperature stages of planet formation. For lava compositions similar to Earth-like mantles or crust, the major atmospheric species are predicted to be Na, O (in both atomic form and molecular \ce{O2}) and \ce{SiO}. \citet{Miguel2011} explored different temperatures (using \texttt{MAGMA}), noting a switch from Na to \ce{SiO} being the dominant species as the temperature is increased.

Currently, the best way of observing the lava atmospheres directly is thought to be using emission spectroscopy in secondary eclipse \citep{Ito2015,zilinskas22}. Transit measurements are currently challenging because of their low atmospheric scale heights. However, the high temperatures of their daysides result in significant emission, particularly in the infrared. \citet{zilinskas22} suggest that {\it JWST} may be able to identify \ce{SiO} and \ce{SiO2} features in atmospheres, which may be indicative of the lava pool's oxidation state, and distinguish atmospheres generated from crust-like and mantle-like pool compositions.

An alternative way of investigating the composition of lava planets is by using the catastrophically evaporating planets. The three confirmed examples (Kepler 1520b, aka KIC 12557548b - \citealt{KIC1255-discov}, KOI-2700b - \citealt{KOI2700b-discov} and K2-22b - \citealt{K2-22b-discov}) were discovered through transit measurements by the {\it Kepler} space telescope. The transit lightcurves appear to be caused by the dust trailing a rocky planet undergoing intense evaporation of its molten surface \citep[see][for a review]{disint18}. In other words, these are lava planets whose silicate atmospheres are escaping and forming dust. Models of the outflows suggest that the planets must have low current masses \citep[$\lesssim 0.1M_\oplus$][]{Perez-Becker13,Booth_disint22} and thus their mass loss is sufficient to destroy the planets within stellar main-sequence lifetimes.

Reaching the goal of using lava planet atmospheres or the dust tails of catastrophically evaporating planets to investigate rocky planet interiors requires understanding how the atmosphere and interior are linked. While the tidally-locked dayside surfaces of lava planets are above the melting point of rocks, neither the resultant pools nor the atmospheres generated are likely to be sufficient to transfer much energy to the nightsides \citep{Kite16}. \citet{CURRY2024} showed that even planets that start fully molten are unlikely to stay molten for more than $\sim$~Myrs due to energy loss from the nightside. This leaves a lava pool that is thin, relative to the radius of the planet, with the rest of the planet solid \citep{Kite16,CURRY2024}.

The lava pool--atmosphere system can evolve significantly, particularly under the influence of mass loss. \citet{Schaefer_fegley2009} considered the effect of fractional vaporisation. The most volatile species in the melt will dominate the atmosphere. Therefore, if the atmosphere is removed, the melt becomes depleted of the most volatile species. The mechanism for the atmospheric removal can either be through day-to-nightside winds \citep{castan-menou11}, or full removal of material from the planet, as is the case for the catastrophically evaporating planets. The latter is believed to be important for low-mass planets, due to low surface gravity, whereas the former dominates for higher-mass planets since it is much less sensitive to planet mass \citep{Kang2021}.

The sequence of elements lost is first Na and K, which are relatively volatile, then Fe, Mg and Si, which essentially leaves behind Al and Ca \citep[a good demonstration of this can be seen in][Fig. 5]{leger2011}. Note that in the melt, all these elements are in the form of oxides, so oxygen is also released into the atmosphere throughout the evolution.

The pool--atmosphere system on these lava pool planets was studied in some detail by \citet{Kite16}. They reached a few notable conclusions. Firstly, if the timescale of circulation in the pools is shorter than that of atmospheric removal then the pools will be uniform. If this is not the case, the pool composition will vary spatially because materials will be removed at different rates from different parts of the pool. The former regime is more common on cooler planets ($\lesssim 2000$~K) as the pool circulation increases less strongly with increasing temperature compared to atmospheric winds. Secondly, using arguments from the theory of horizontal convection they argue that the lava pools may be as shallow as only a few 10s of m deep (see \autoref{sec: pool depth}). Thirdly, fractional evaporation may cause the density of the melt left behind to increase or decrease. If the density decreases the residual material will create a buoyant lid and continue to lose its volatiles, whereas if it increases the residue will sink and the composition close to the surface of the pool will be reset with unevolved material. They also speculate that, as material is added into the pool from the mantle below, it should be possible to reach a steady state where the material evaporated has the same composition as that melted into the pool, although they do not model the process explicitly. This scenario implies a steady state pool composition, which is {\it not} the same as the composition of the material entering and leaving the pool.  

In this work, we develop a simple model of the chemistry of the lava pool--atmosphere system, focussing on the effect of mass loss. The model advances the works of \citet{Schaefer_fegley2009} and \citet{Kite16}, with the inclusion of melting into the pool, which enables the steady state situation proposed by \citet{Kite16} to be reached. Furthermore, we consider both catastrophically evaporating planets and more massive planets where the evolution is driven by day-to-night-side winds. We begin, in \autoref{sec: overview}, with an overview of the physical situation we are modelling and present a simplified model, which captures many features of the full model. Then, in \autoref{ch:Mpool sec:model}, we describe the operation of the full model and present its results in \autoref{ch: Mpool sec: pool_results}. In \autoref{ch:Mpool sec: likelihood of evolved}, we assess the likelihood of pools reaching highly evolved states. We discuss some of the model's caveats and wider implications in \autoref{ch:Mpool sec:discuss} and summarise our conclusions in \autoref{ch:Mpool sec:summary}.

\section{Problem overview}\label{sec: overview}
\begin{figure}
    \centering
    \includegraphics[width=\linewidth]{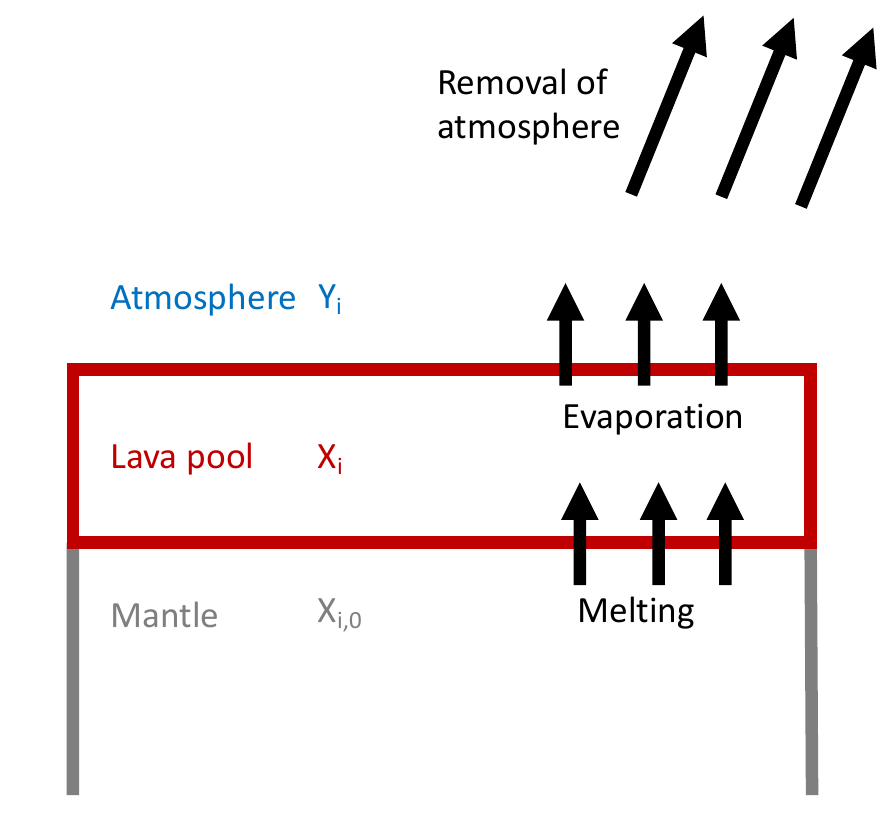}
    \caption[Schematic of lava pool chemical evolution model]{Schematic of the chemical evolution model described in \autoref{sec: overview} and \autoref{ch:Mpool sec:model}. $Y_i$, $X_i$ and $X_{i,m}$ are the concentrations of element $i$ in the atmosphere, pool and mantle, respectively. We consider material flowing in and out of a lava pool of fixed mass, by evaporation and melting, which are marked in black. The general evolution of the pool composition is governed by \autoref{ch:Mpool  eq: general simp model}. The atmosphere can be removed entirely from the planet, via atmospheric escape, or to the nightside, via day-to-nightside winds.}
    \label{fig: pool_atm_schem}
\end{figure}

The basic physical setup we will model in this work is shown in \autoref{fig: pool_atm_schem}. There is a fully mixed molten region which generates vapour which is continually removed by some process (either removal from the planet or to the nightside). Following \citet{Schaefer_fegley2009}, we assume that the elements in the vapour are purely those which make up the bulk of the molten region, i.e., any volatile species (e.g., C, N, S) have already been removed from the planet by escape or were never incorporated in the planet in the first place. As material is removed, new material is also continually melted from the underlying mantle into the base of the pool. The composition of melt added to the pool will depend on the mantle composition and also the conditions at the mantle--pool interface, but we will just consider this as an input parameter and consider a few different compositions in \autoref{ch: Mpool sec: pool_results}. The composition of the material removed will depend on two things. Firstly, the volatility of species will determine what enters the atmosphere from the pool. Secondly, different species may escape from the atmosphere at different rates, which is termed `atmospheric fractionation' \citep{ZAHNLE1986}. In the following subsection, we assess the likelihood of atmospheric fractionation. In \autoref{sec: general eqns}, we lay out the general equations that govern the chemical evolution of the pool. Then we describe a simplified model of the physical scenario, in \autoref{ch:Mpool sec:simplified}, before describing the full model in \ref{ch:Mpool sec:model}.

\subsection{Atmospheric fractionation due to escape} \label{ch:Mpool sec: fractionation escape}
As species escape from multi-component atmospheres, they can escape at different rates according to their masses. In this section, we briefly assess the likelihood of fractionation for catastrophically evaporating planets' atmospheres, where full escape from the planet occurs. This analysis will not apply to higher mass planets for which loss from the lava pool is instead dominated by day-to-nightside winds \citep{Kang2021}. In that case, fractionation may occur differently, but it is likely that the condensation back into the pool will instead be the dominant process \citep{Kite16}. We therefore do not consider the fractionation in those cases any further.

Atmospheric fractionation has long been known to be important in the evolution of solar system atmospheres \citep[e.g.,][]{Hunten1987,Zahnle1990,Lammer2008}, and is now of interest for exoplanet atmospheres \citep[e.g.,][]{lugerBarnes2015,AIOLOS2023}. Roughly speaking, low number density weakly escaping atmospheres undergo strong fractionation, whereas for dense and strongly escaping atmospheres species can effectively drag on each other, meaning the escape becomes coupled and fractionation does not occur.

For an escaping atmosphere consisting of multiple species, one can define the escape factor of a species $i$ as the ratio of the number of particles of $i$ that escape to that of the most strongly escaping species. This is given by \citep{Yung1988,Zahnle1990}
\begin{equation}
    x_i = \frac{\Phi_i}{F_i \Phi_1} \; , \label{ch:Mpool eq: def escape factor} 
\end{equation}
where $\Phi$ is the particle flux, the subscript `$1$' denotes the most vigorously escaping species and $F_i$ is the mixing ratio at the base of the outflow of the species $i$ relative to species $1$ (i.e., $F_i = n_i/n_1$, where $n$ is the number density.) If $x_i = 1$ then $i$ is fully coupled to the outflow, meaning it escapes with the same mixing ratio as it has in the lower atmosphere. If $x_i = 0$ the species does not escape at all.

In general, the full multi-species fluid equations need to be calculated to find escape factors. However, under certain conditions, analytical estimates can be made. One such estimate, the derivation of which can be found in \citet{Zahnle1990}, applies to atmospheres made of two escaping species: 
\begin{equation}
     x_2 = 1 - \frac{GM(m_2-m_1)b_{12}}{r_0^2k_BT(1+F_2)\Phi_1} \; .\label{ch:Mpool eq: two-species escape factor}
\end{equation}
Here $G$ and $k_B$ are Newton's gravitational constant and the Boltzmann constant; $M$ is the planet mass; $r_0$ is radius at the base of the outflow; $m_i$ is the mass of species $i$; $T$ is the temperature of the outflow and $b_{ij}$ is the \textit{binary diffusion coefficient}, a measure of how well a species diffuses through another species.

The atmospheres considered in this work (see \autoref{ch: Mpool sec: pool_results}) have more than one species. However, this two-species escape factor is still a useful measure of how strongly one species will drag on another, which allows us to broadly assess whether or not the atmospheres are well coupled. The calculations are simpler if the escape is coupled, so we will attempt to show that the escape is indeed coupled under the conditions we are interested in.  Therefore, when calculating escape factors ($x_2$ in \autoref{ch:Mpool eq: two-species escape factor}), we set $F_2=0$, which gives the highest chance of a species being uncoupled as $x_2$ is minimised\footnote{As it happens it also gives the limit when species 2 is trace, which is valid regardless of how many trace species there are, as long as there is only one major species. However, this is not very relevant for our calculations.}.

\begin{figure}
    \centering
    \includegraphics[width = \linewidth]{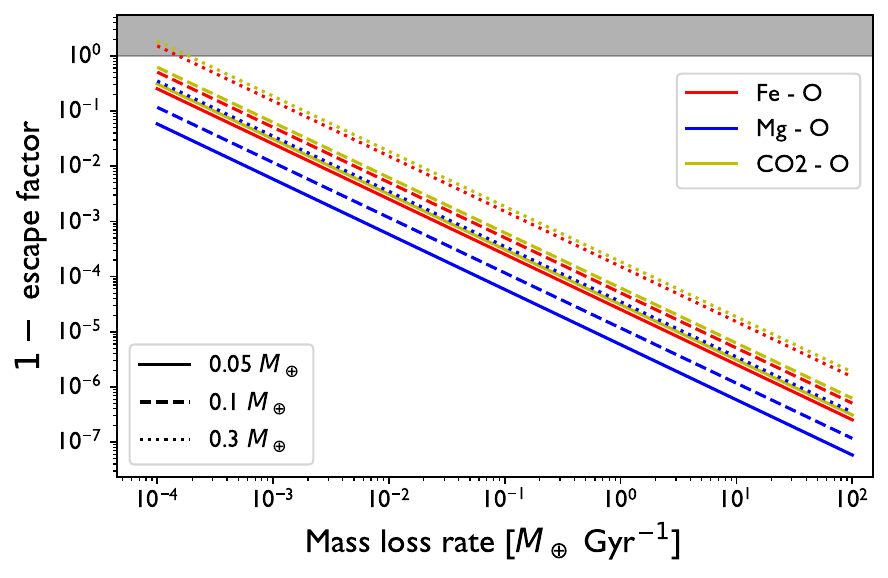}
    \caption[Escape factors]{1 - escape factor ($x_2$ calculated using \autoref{ch:Mpool eq: two-species escape factor}) for different species combinations, mass-loss rates and planet masses. Values close to 1 mean the atmosphere fractionates strongly and close to 0 mean it is well coupled (the escape factor is close to 1). For all atmospheres, the temperature is assumed to be 2000 K. Values in the grey $x_i < 0$ region are unphysical (see text).}
    \label{fig: escape factors}
\end{figure}

In \autoref{fig: escape factors}, we plot the quantity $(1-x_2)$, where $x_2$ is the escape factor given by \autoref{ch:Mpool eq: two-species escape factor} (with $F_2=0$), for a few relevant species combinations. This quantity is the second term on the right-hand side of \autoref{ch:Mpool eq: two-species escape factor} and is a more convenient quantity to plot because $x_2$ is very close to 1 for most values. Lower values of $(1-x_2)$ correspond to more coupled atmospheres. The calculation of the theoretical binary diffusion coefficients $b_{ij}$ is described in Appendix \ref{app: binary diff}, and some atomic parameters are shown in \autoref{ch: Mpool tab: gas diff params}. $r_0$ is assumed to be the planet's surface and is calculated using the mass-radius relationship from \citet{Fortney07} with a core mass fraction of 0.3. The lower the value on the $y$\nobreakdash-axis, the better coupled the two species are. These values are independent of the number density in the atmosphere since this is entirely taken into account through the mass-loss rate. The calculations shown are for 2000~K, although it turns out that the temperature dependence is very small (see Appendix \ref{app: binary diff}). 

At very low mass loss rates (1-$x_i$)$>1$ (grey region in \autoref{fig: escape factors}), which implies a negative escape factor, which is unphysical. This is simply because \autoref{ch:Mpool eq: two-species escape factor} is no longer valid. Its derivation assumes that all species are escaping. However, if the lighter species' mass loss rate is reduced enough, it will no longer be able to drag the heavier species, and that species will not escape. 

The choice of iron and magnesium coupling to oxygen is inspired by the types of species produced in lava pool atmospheres (see \autoref{ch: Mpool sec: pool_results}). We also show O-CO$_{2}$, a molecule combination for which the diffusion coefficient can be experimentally measured, to demonstrate that the theoretically derived values (Fe-O and Mg-O) are of a reasonable order of magnitude (\ce{CO2} has a mass between atomic \ce{Mg} and \ce{Fe}). For $b_{O-CO_2}$, we use the experimentally derived formula given in \citet{ZAHNLE1986}.

The catastrophically evaporating planets are believed to have mass-loss rates exceeding 0.1 $M_\oplus$ Gyr$^{-1}$ \citep{Perez-Becker13,Booth_disint22}, which, according to \autoref{fig: escape factors} is in the well-coupled regime. Earlier in those planet's lifetimes, they will have had lower mass-loss rates. That being said, to get to their observable stage, they still must have started with reasonably high mass loss rates \citep[$\gtrsim~10^{-2}$~$M_\oplus$~Gyr$^{-1}$,][their Figure 6]{CURRY2024}. Therefore, it is reasonable to assume that the catastrophically evaporating planets have undergone coupled mass loss. \autoref{fig: escape factors} does demonstrate that there is a regime where fractionation due to escape may be important, for high-mass planets and low mass-loss rates. However, as shall be discussed in \autoref{ch:Mpool sec: likelihood of evolved}, in that regime, day-to-night atmospheric transport is likely to be more important, and so this fractionation is not as relevant for atmospheric evolution.

Therefore, we believe that it is justified to henceforth assume that the atmospheric loss is coupled and does not undergo mass fractionation.

\subsection{General equations of pool evolution}\label{sec: general eqns}
In general, the time evolution of the number of particles of any species (atoms if tracking elements, or cations if tracking the oxides) in the pool is governed by the differential equation
\begin{equation}
    \diff{N_i}{t} = -a_i(X_i,X_j...) + b(N_i,N_j...)X_{i,m} \; ,\label{ch:Mpool  eq: general simp model}
\end{equation}
where $X_i$ is the concentration of species $i$, so $N_i = X_i N$ where $N$ is the total number of particles in the pool. $N$, $X_i$ and $N_i$ are all functions of time $t$. Here $a_i$ controls the removal of material due to its evaporation, and $b$ controls the replenishment of material with composition $X_{i,m}$, where $m$ stands for mantle. In our evolutionary runs in this and in later sections, we will also assume that $X_{i,m}$ is the initial composition of the pool because the pool is formed directly from the melted mantle, although it does not necessarily need to be. 

This equation clearly has the potential to reach a steady state, where the two terms on the right-hand side are equal resulting in a constant composition and mass ($\mathrm{d}N_i/\mathrm{d}t = 0$ for all $i$). The physical interpretation of the balance of the right-hand side terms is that the material entering and leaving the pool have the same composition. This composition 
is not the same as the constant pool composition ($N_i$). We will find that this steady state is indeed reached under the conditions that we solve this equation.

It is useful to change the independent variable of this ODE from time to mass removed to the atmosphere, which enables us to conduct an analysis which just depends on the mass removed from the system, regardless of how fast it is occurring, and is thus applicable in more situations. To do so, we note that the rate of removal of mass from the pool is given by 
\begin{equation}
    \diff{m_\text{out}}{t} = \sum_j a_j(X_j,...) \, m_j \; , \label{ch:mpool eq:dmdt}
\end{equation}
where $m_j$ is the molecular mass of the species $j$.
Equations \ref{ch:Mpool  eq: general simp model} and \ref{ch:mpool eq:dmdt} can then be combined to give
\begin{equation}
    \diff{N_i}{m_\text{out}} = \frac{-a_i(X_i, ...) + b(N_i,...)\,X_{i,m}}{\sum_j a_j(X_j...)\, m_j} \; . \label{ch:Mpool eq: dN_i_dm}
\end{equation}
For much of this paper, we will essentially be solving this equation. 

Both in this simplified model in the next subsection and in the full model in \autoref{ch:Mpool sec:model}, we will assume that the pool has {\it constant mass}. The depth (and thus mass evolution) of an irradiation-induced magma pool is not well constrained, as we discuss in \autoref{ch:Mpool sec:non const mass}, so constant mass is ultimately a simplifying assumption. However, in \autoref{ch:Mpool sec:non const mass}, we also argue that the depth of the pool is unlikely to alter much as a planet loses mass and its composition changes, so the assumption is reasonable.

The additional advantage of a constant pool mass is that it makes the analysis fully independent of the actual pool mass and rate of atmospheric loss. The reason is that the mass of material added to the pool automatically equals that removed. Thus the rate of removal does not matter in terms of determining the composition because if one rate speeds up, so does the other. The only determinant of the instantaneous composition of the pool is the initial composition and how much mass has been removed to the atmosphere. However, as the amount of a species moved to the atmosphere depends on the concentration of the species, not the total mass of the species, all that really matters is the amount of mass removed \textit{relative} to the total pool mass. Indeed the units we shall use to study the evolution will be mass removed in terms of the number of pool masses.
In \autoref{ch:Mpool sec: likelihood of evolved}, we actually address the quantities of pool mass and mass loss rate, allowing the results to be put into the context of the exoplanet population. 

By considering \autoref{ch:Mpool  eq: general simp model} and \autoref{ch:mpool eq:dmdt}, one can deduce that to maintain a constant pool mass requires 
\begin{equation}
    \sum_j a_j(X_j,...)m_j =  \sum_j b(X_i,X_j...) m_j X_{j,0} \; . \label{eq: b constraint}
\end{equation}
This constrains the value of $b$.

\subsection{Simplified pool model}\label{ch:Mpool sec:simplified}

Before describing our full model in \autoref{ch:Mpool sec:model}, we first demonstrate a simplified version of it. As we will see, this toy model produces most of the same features as the full model and thus is helpful for understanding it.

The primary determinant of the amount of a species removed is the amount of that species in the melt, so for the simplified model we drop all other dependence in the first term. We further assume that the dependence on melt concentration is linear, i.e., $a_i(X_i,X_j,...) \rightarrow \hat{a}_iX_i$, where $\hat{a}_i$ is a species-dependent constant encoding volatility. In general, this may not be the case (see \autoref{eq: general partial pressue}), but we shall see that the main features are reproduced regardless.

For the simplified model, we take the atomic masses ($m_i$) of all species to be the same and equal to 1. This means that the assumption of constant pool mass results in a constant total number of particles in the pool $N$. $N$ can therefore factorise out of the time derivative in \autoref{ch:Mpool eq: dN_i_dm}.
The pool mass may, without loss of generality, be set to 1, which simply defines the units we use for the amount of mass removed.

These two simplifications mean that \autoref{ch:Mpool eq: dN_i_dm} can be rewritten as
\begin{equation}
    \diff{X_i}{m_\text{out}} = \frac{-\hat{a}_i X_i + b(X_i,X_j...)X_{i,m}}{\sum_j \hat{a}_j X_j } \; , \label{ch:Mpool eq: dx_i_dm}
\end{equation}
where the constant total number of particles $N$ has been absorbed into the coefficients $\hat{a}_i$ and $b$. Using \autoref{eq: b constraint},
\begin{equation}
    b(X_i,X_j...) = \frac{\sum_j \hat{a}_j X_j}{\sum_j X_{j,0}} \; . \label{eq: b simplified}
\end{equation}

\begin{figure}
    \centering
    \includegraphics[width=0.9\linewidth]{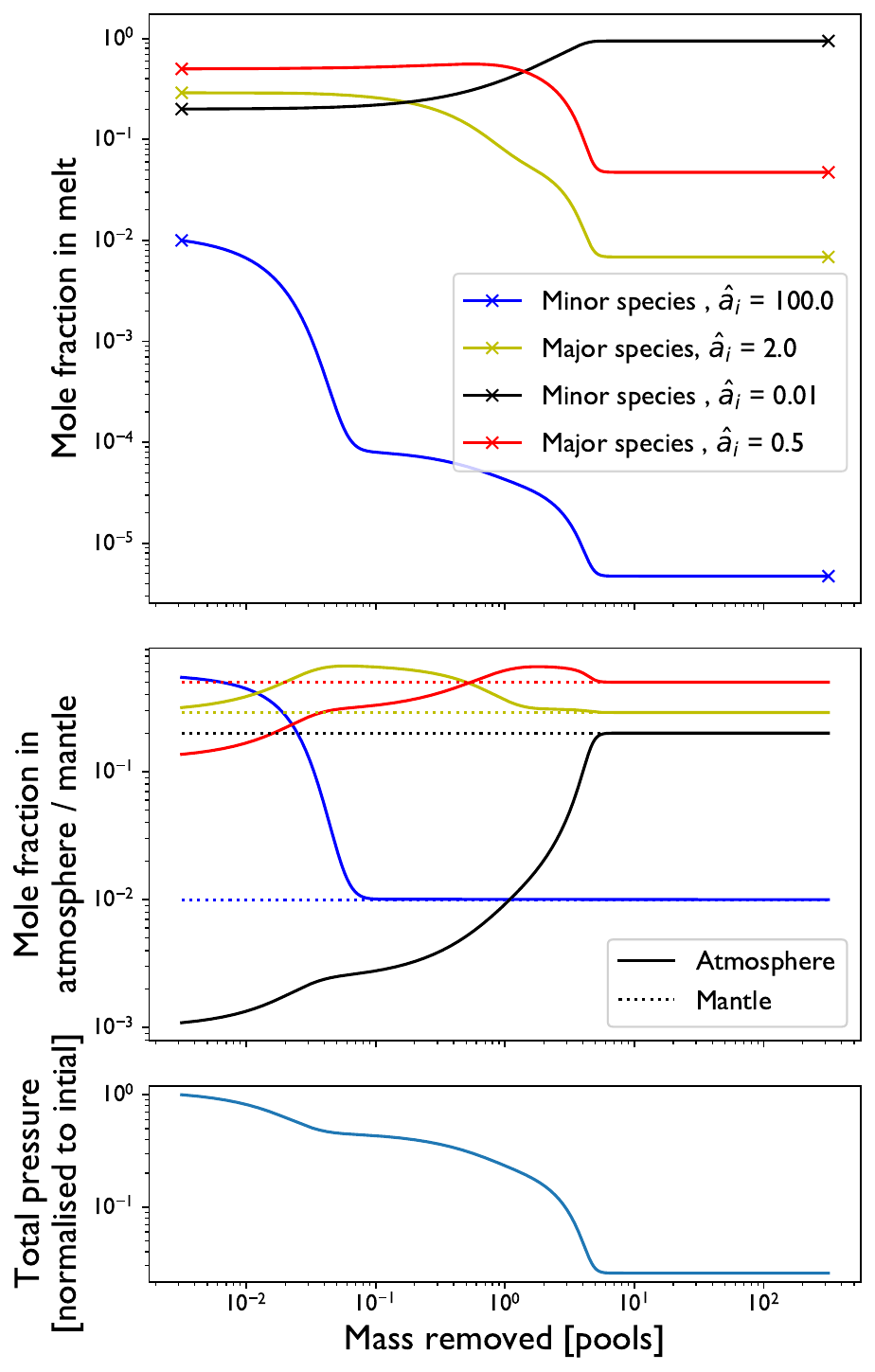}
    \caption[Simplified lava pool model]{Abundances of melt and gas species and the total pressure at the base of the atmosphere in our simplified pool model (\autoref{ch:Mpool sec:simplified}.) The coefficients $\hat{a}_i$ correspond to relative volatilities (with a larger value denoting a higher volatility).}
    \label{fig: simplified_pool}
\end{figure}

To see how this simple system evolves, we integrated \autoref{ch:Mpool eq: dx_i_dm} using \texttt{scipy odeint} for a four-species system, the result of which is shown in \autoref{fig: simplified_pool}. \texttt{Odeint} uses the LSODA method \citep{hindmarsh83}, which dynamically chooses between stiff and non-stiff methods with adaptive step sizes. We use the default relative and absolute tolerances of $2^{-26}$. For the example shown here, we picked relative volatilities (shown in the key) and a mantle composition so as to produce an evolution that looks similar to that of Bulk Silicate Earth (see \S\ref{ch:Mpool sec:BSE} and \autoref{fig: BSE evolution}, below). The starting composition of the pool is the same as that melted from the mantle ($X_{i,m}$). It is important to note that in this and following plots the $x$-axis is mass lost \textit{not} time. How mass loss maps onto time evolution is addressed in \S\ref{ch:Mpool sec: likelihood of evolved}. As stated above, \autoref{ch:Mpool  eq: general simp model} clearly has a steady state solution, and, for the simple model (\autoref{ch:Mpool eq: dx_i_dm}), one can predict the melt composition that satisfies this steady state because of the linear dependence of removal to the atmosphere on concentration. This steady state composition is given by
\begin{equation}
    X_{i,f} = \frac{b X_{i,m}}{\hat{a}_i} \; .
\end{equation}
In practice, to work out $X_{i,f}$ one does not need to use \autoref{eq: b simplified} to calculate $b$; one can instead use the fact that, by their definition as concentrations, the $X_{i,f}$, must sum to 1. The predicted steady state solutions are shown as crosses in the upper panel of \autoref{fig: simplified_pool}.

\autoref{fig: simplified_pool} demonstrates that the pool composition does indeed reach an equilibrium, with a final composition as predicted (see upper panel). It does so after $\sim10$ pools worth of mass have been removed. There is some intuitive sense to this value. The final state will be one when the most volatile species have been removed and so the least volatile species dominates. The least volatile species makes up around 10\% of the initial pool mass in this example, so it takes around 10 reprocessings for it to become the dominant species.

The most volatile species (blue and yellow) have monotonically decreasing abundances in the melt, although they flatten out for periods in the evolution. This is because it is possible to reach a pseudo-steady state where $b$ in \autoref{ch:Mpool eq: dx_i_dm} changes slowly. The red species, which has an intermediate volatility, initially increases in abundance in the melt as the more volatile species are removed. However, it later drops off as it is removed faster than the most refractory species (black). When the steady state is reached, the melt abundances are ordered from least to most volatile (with the least volatile being most abundant), as one would expect.

The middle panel of \autoref{fig: simplified_pool} shows the evolution of the abundances in the vapour. Changes in the vapour abundances track those in the melt. It is interesting to note that the least abundant species in the melt (blue) is the most abundant in the vapour at the start because it is so volatile. However, when the atmosphere is removed, this species is removed and replaced by further vaporisation from the melt. This means it quickly becomes depleted in the melt, and so also the atmosphere. The final composition in the vapour perfectly matches that in the mantle, because this must be the case to maintain a steady state of material entering and leaving the melt (and thus also a steady state in the melt composition itself).  It is helpful to note when interpreting the middle panel of \autoref{fig: simplified_pool} and future similar figures that, since it shows the fraction in the atmosphere (and not the total number of moles), some species may have a continually dropping partial pressure ($P_i \equiv Y_i P$, where $P$ is the total pressure) and yet increase as a fraction of the atmosphere because other species' pressures are dropping faster. This is the case for the yellow species, whose abundance in the melt is continually dropping so its partial pressure must be too, yet its atmospheric fraction increases.

The bottom panel of \autoref{fig: simplified_pool} shows the evolution of the total pressure. The pressure in the simplified model has arbitrary units so we simply normalise it to the initial value. The total pressure monotonically decreases, which it should as volatile species will result in a higher pressure atmosphere, and these dominate initially but are removed. There are two regions where the pressure levels out, similar to the concentrations in the top panel. The reason is naturally connected. The total pressure in the atmosphere is mostly controlled by the partial pressure of the dominant vapour species and this is controlled by its concentration in the melt. 

\section{Chemical evolution model}\label{ch:Mpool sec:model}
In this section, we move on from the simplified model in the previous section and describe a more complex model that includes actual lava pool atmospheric chemistry calculations. The simplified model only used dummy species with parameterised volatilities, which all have the same atomic mass. In our full model, we calculate the composition of an atmosphere in equilibrium with a lava pool of a certain composition in order to find what species are removed as the atmosphere escapes. As with the simplified model above, we take the pool to have a fixed mass. 

The procedure to evolve the lava pool's composition forward in time is as follows. Firstly, we calculate the composition of an atmosphere that exists in chemical equilibrium with a lava pool of a given composition, which will be described in \autoref{ch:Mpool sec: calc atm comp}. The elements in this atmosphere have come from the pool, so we remove a small amount of material from the pool with the same composition as the vapour (`evaporation' arrow in \autoref{fig: pool_atm_schem}). Removal of material of this composition assumes that, as the atmosphere moves away from the pool, there is no further fractionation of species, which we justified in \autoref{ch:Mpool sec: fractionation escape}. The same mass of material that was removed from the pool is then added from the mantle, so that the pool's mass remains the same. The procedure is then repeated. Mathematically this corresponds to the number of particles in the pool of element $i$ on the $n+1^\text{th}$ step being
\begin{equation}
     N_i^{n+1} = N_i^{n} - c Y_i^n + b X_{i,m} \; , \label{ch:Mpool eq: next step Ni}
\end{equation}
where $Y_i^n$ is the concentration of the element in the vapour, $c$ is a constant that is the same for all elements, as is $b$, which is as defined in \autoref{sec: general eqns}. This is the discretised form of \autoref{ch:Mpool  eq: general simp model} but identifying $a_i(X_i,X_j...) \,\propto\, Y_i$. One can write this equation instead in terms of the evolution of concentrations $X_i$, using the definition of concentration ($N_i = NX_i$):
\begin{equation}
     X_i^{n+1} = \frac{X_i^{n} - c Y_i^n + b X_{i,m}}{N^{n+1}/N^n} \; , \label{ch:Mpool eq: next step Xi}
\end{equation}
where the constants $b$ and $c$ have absorbed a factor of $N^n$.

In order for the mass of the pool to be constant
\begin{equation}
    b = \frac{c \sum_i  Y_i^n m_i}{\sum_i  X_{i,m} m_i } \; , \label{ch:Mpool eq:b_same_pool_mass}
\end{equation}
which is derived in the same way as the condition in \autoref{eq: b constraint}. In the next subsection, we describe how we calculate $Y_i$, the atmospheric composition. In \autoref{ch:Mpool sec:step_control}, we describe details of the numerical procedure, which determines~$c$.

\subsection{Atmospheric composition}\label{ch:Mpool sec: calc atm comp}
At a given temperature and pressure, lava will reach a chemical equilibrium with an atmosphere produced above it. Assuming the chemical timescales are short enough relative to the timescale of atmospheric removal, one may assume the composition of an atmosphere above a magma pool has this equilibrium composition. Since the atmospheres we consider have very high temperatures, this assumption of chemical equilibrium is justified. As the atmosphere is removed, the magma pool's composition will change, and thus so will the atmosphere's, but instantaneously one can consider the equilibrium to hold.

To determine this composition we used the open source code \texttt{LavAtmos} \citep{LavAtmos}. Vaporisation of lava occurs through various chemical reactions of the general form
\begin{equation}
    \ce{ p M_{q}O_{r}(l) + s O2(g) <=> M_{p q}O_{r+2s}(g)} \; , \label{eq: gen chem eqn}
\end{equation}
where M is a cation, $q$ and $r$ are the number of cations and oxygen atoms in the liquid oxide, and $p$ and $s$ are stoichiometric coefficients ($s$ can be positive or negative). `l' and `g' subscripts denote liquid and gas phases respectively. \texttt{LavAtmos} takes a lava composition and finds the chemical activity~$a_i$ of its constituent oxides using the thermodynamic model \texttt{MELTS} \citep{MELTS1,MELTS2}. It then uses data from the \textit{JANAF} database \citep{NIST-JANAF}, to find the equilibrium constants $K(P,T)$ for a collection of reactions allowing the partial pressures of each gas to be calculated as
\begin{equation}
   P_{M_{pq}O_{r+2s}} = K(P,T) \left(a_{M_qO_r}\right)^p \left(f_{\ce{O2}}\right)^s \; , \label{eq: general partial pressue}
\end{equation}
where $f_{\ce{O2}}$ is the oxygen fugacity. 

To find a unique solution to the chemical equilibrium, the oxygen fugacity $f_{\ce{O2}}$ (or partial pressure of \ce{O2(g)} if the gas is assumed to be ideal), must be constrained. \texttt{LavAtmos} works out $f_{\ce{O2}}$ through the `law of mass action', which simply means that the ratio of cations to oxygen in the atmosphere is the same as in the lava. In other words, all the \ce{O2(g)} in the atmosphere comes from the liquid oxides. This means that the code has to iterate in order to find a solution such that this is true. 

The other recent open source code \texttt{VapoRock} \citep{VapoRock}, which works almost identically, instead keeps $f_{\ce{O2}}$ as a free parameter. Their motivation for this is that the oxygen fugacity has a very large effect on atmospheric compositions \citep[e.g.,][]{Kasting1993} and its value depends strongly on the composition and mineralogy of the material the atmosphere is in contact with \citep{Frost1991,annurev-mantle-fugacity,Guimond_2023}. It can thus act as a simple proxy for different compositions defined by how oxidising they are, without having to consider in detail what the compositions actually are. This approach, where fugacity is effectively considered a property of the material independent of composition, is used extensively in the literature to study outgassing from magma oceans or volcanism. For the evolutionary calculation we are performing, however, the oxidising ability of the lava should evolve as its composition does. Therefore, a constant $f_{\ce{O2}}$, independent of the composition, would not be appropriate. The `law of mass action' is a self-consistent way of allowing its value to change. In principle, to investigate the effect of more reducing or oxidising lavas the composition can be changed.

\texttt{LavAtmos} requires an input pressure, as the equilibrium constants are functions of pressure (as written explicitly in \autoref{eq: general partial pressue}). Technically, composition solutions will only be self-consistent if the input pressure matches that of the generated atmosphere. This requires iteration of the chemistry calculations with a continually updated input pressure. However, the constants are very weak functions of pressure \citep{LavAtmos} for the pressure ranges of lava pool atmospheres ($\lesssim10$ bar), so the resulting atmospheres are not greatly affected, and the extra iterations are unnecessary. In our evolutionary calculations, we use the pressure of the atmosphere in the previous step as the pressure for the equilibrium constants, so if evolutionary steps are small, the calculations should be close to being self-consistent. Furthermore, we are expecting to evolve to a steady state, as was found in \autoref{ch:Mpool sec:simplified}, at which point this time-lagging will not matter anyway.

An assumption that is implicitly made by using \texttt{LavAtmos} in this way to calculate lava pool atmospheres is that the mass of the pool reservoir is much greater than that of the atmosphere. Were this not the case, the formation of an atmosphere would alter the pool's composition (even without any atmospheric removal) and so the equilibrium between atmosphere and lava would no longer be valid. Further iterations of the solution would be required to find the true equilibrium. Clearly, figuring out whether this is needed requires knowing the mass of the pool and atmosphere, so, for now, we will assume that the pool's mass reservoir dominates, and show {\it a~posteriori} that in most cases it does (\autoref{ch:Mpool sec:atm mass}).

\subsection{Evolutionary step control}\label{ch:Mpool sec:step_control}
To maintain accuracy, the composition of the lava should not change too much over each evolutionary step, thus approximating a continuous process. As noted at the start of \autoref{ch:Mpool sec:model}, the amount of an element removed from the melt at each step is proportional to its concentration in the vapour, $Y_{i}$. To ensure the compositional changes are not too high, we select the element with the highest $Y_{i}/X_i$ ratio (as a reminder, $X_i$ is the concentration of $i$ in the melt), which we denote as element $v$ (for {\it volatile}). We then remove some $\Delta X_{v} = \xi X_{v}$, where $\xi$ is some small value we choose. All other elements $i$ are then removed, with the amount removed scaled to their vapour elemental ratios:
\begin{equation}
    \Delta X_i = c Y_i \text{ , where } c \equiv \frac{\xi X_{v} }{Y_{v}} \; . \label{ch:Mpool eq:scaling_vap_to_removed}
\end{equation}
As the element $v$'s fractional amount in the melt changes the most, this ensures that all other species are changed by an even smaller amount.

The concentrations $X_i$ of volatile elements will decrease steeply as they are removed from the melt. This means that to maintain accuracy ever smaller steps $\Delta X_i$ must be taken ($c$ in \autoref{ch:Mpool eq:scaling_vap_to_removed} becomes small\footnote{One might note that $c \, \propto \, X_v/Y_v$ and that $Y_v$ will also decrease, meaning $c$ need not necessarily be small. However, as the elements are volatile $Y_v$ remains substantial even with very low $X_v$.}). Consequently, computing the evolution becomes very slow. Since the pool is continually replenished by material melted from the mantle, one cannot simply remove these elements. However, one can take larger steps, and so speed up computation, by exploiting the fact that a steady state should be reached where the amount of an element being lost from the pool to the atmosphere is the same as that being added through melting. 

The concentration of an element in the $n+1^\text{th}$ step is given by \autoref{ch:Mpool eq: next step Xi}. If, for the element with the highest $Y_v/X_v$ (denoted with $v$ subscript again),
\begin{equation}
  |X_v^{n+1} - X_v^{n}|/X_v^{n} < \varepsilon \; , \label{ch:Mpool eq:pseudo_eqm_condition}
\end{equation}
where $\varepsilon$ is some small threshold, then we deem the element to be in a pseudo-steady state and set $X_v^{n+1} = X_v^{n}$. The element with the next highest $Y_{i}/X_i$ is then used to determine the amount of material removed in that step instead. If it also satisfies the pseudo-steady state condition, it is also set to have the same number fraction as in the previous step, and the next element is considered etc. As this should move to elements with higher $X_i$, it will allow a larger step to be taken. We call this a pseudo-steady state because the functions $Y_{i}(X_i)$ (calculated using \texttt{LavAtmos}, \autoref{ch:Mpool sec: calc atm comp}) and $b(N_i,N_j)$ are functions of the overall composition, thus the steady state will only be true for small changes in the total composition. After taking a longer step, or a number of steps, the condition of $|X_i^{n+1} - X_i^{n}|/X_i^{n}$ being small enough may no longer be satisfied. At this point, smaller steps must be taken again, until a new pseudo-steady state is reached. This means it is necessary to go through the process of checking for each species at every step, even if some were in a pseudo-steady state at the previous step. Furthermore, so that the next step remains close to the pseudo-steady state, we also impose a restriction on the absolute amount of an element that can be removed, as well as fractional.

For the evolutionary runs in \autoref{ch: Mpool sec: pool_results}, $\xi=0.05$, $\varepsilon = \num{2e-4}$ (other than the additional models in Appendix \ref{ch:Mpool sec:limit to eqm}) and the absolute concentration of any element is not allowed to change by more than $10^{-5}$. Our results are robust to changes in the exact values of these model parameters. A particular case of changing the step size (using $\varepsilon$) and converging to the same solution is demonstrated in Appendix \ref{ch:Mpool sec:limit to eqm}.

\section{Results of chemical evolution model} \label{ch: Mpool sec: pool_results}
\begin{figure}
    \centering
     \includegraphics[width=\linewidth]{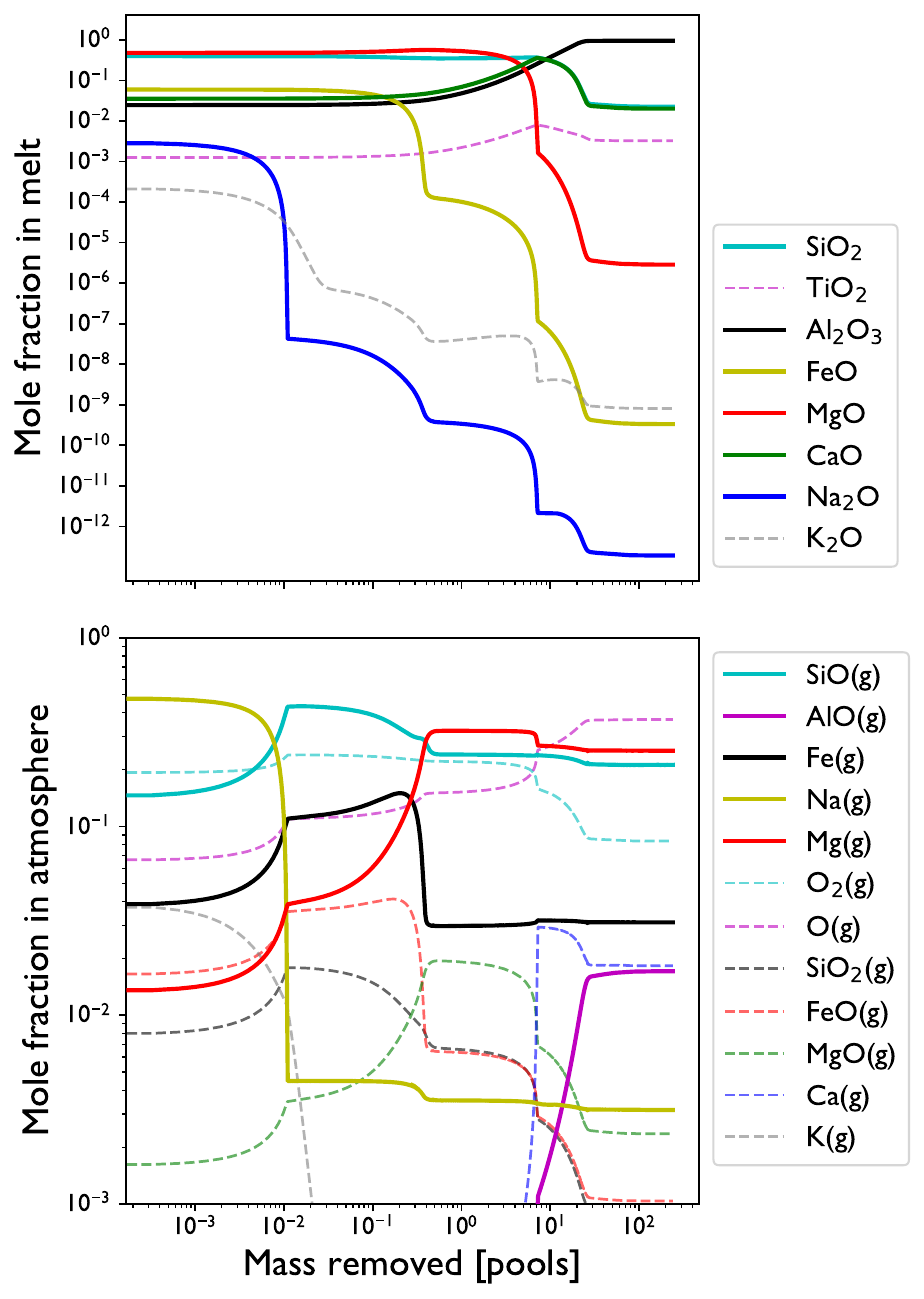}
    \caption{Evolution of lava pool and vapour compositions for a mantle with {\it BSE} composition at a temperature of 2600 K. All melt species are shown. Only vapour species that have a mole fraction above 1\% at some point in the atmosphere's evolution are shown. As in \autoref{eq: gen chem eqn}, `g' denotes gas species. We deliberately highlight certain species as opaque solid lines to make the evolutionary sequences described in the text easier to see. There is otherwise no difference indicated by the formatting of the lines.}
    \label{fig: BSE evolution}
\end{figure}

\begin{figure}
    \centering
    \includegraphics[width=\linewidth]{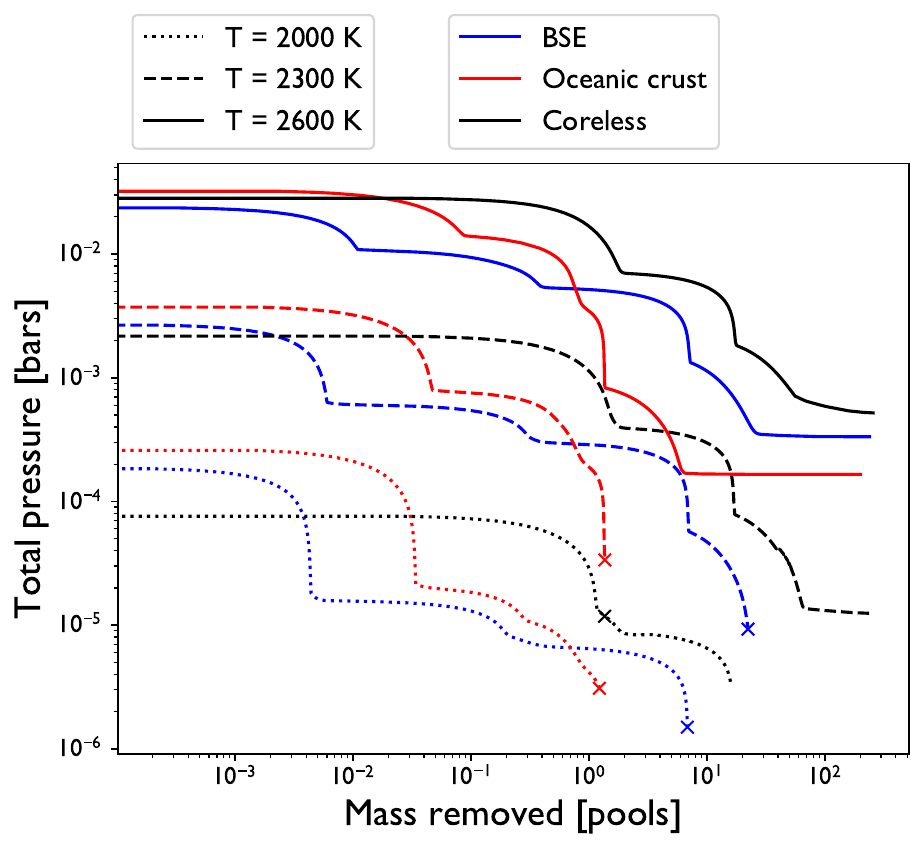}
    \caption[Evolution of total pressures of lava atmospheres]{Evolution of the total pressures of equilibrium lava atmospheres produced by a mantle with different compositions and at different temperatures. Crosses mark when and if the pool becomes partially solid.}
    \label{fig: total_pressures}
\end{figure}

\begin{figure}
    \centering
   \includegraphics[width=\linewidth]{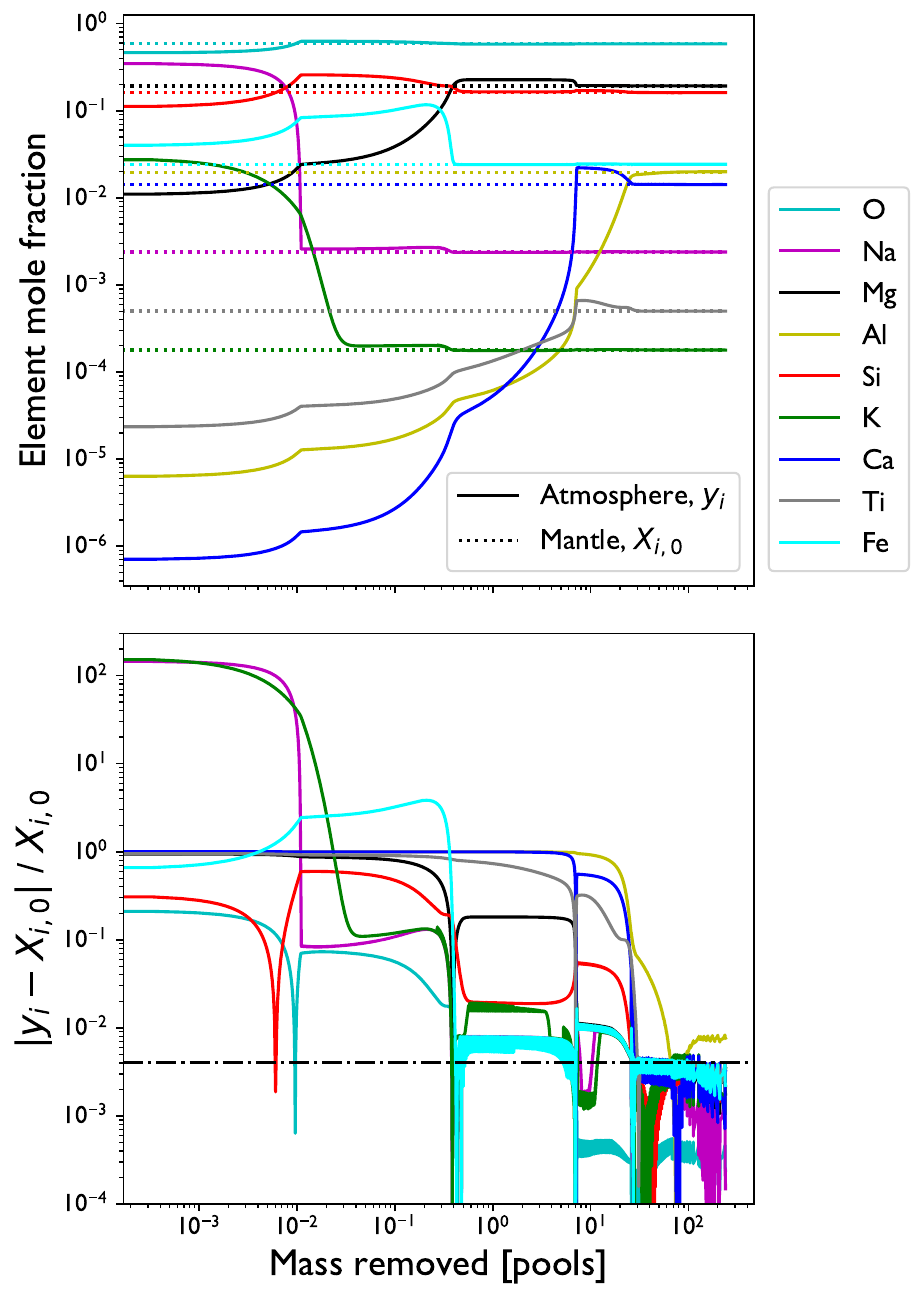}
    \caption[Convergence to an equilibrium state of the lava pool--atmosphere system]{Demonstration of the convergence to an equilibrium composition for a lava pool--atmosphere system for a mantle with {\it BSE} composition at a temperature of 2600 K (evolution shown in \autoref{fig: BSE evolution}.) The top panel shows the element mole fractions for all elements in the atmosphere and the mantle, and the bottom panel shows the fractional difference between these values for each element. The dot-dashed line shows the predicted threshold for convergence of elements other than \ce{Al} (see Appendix \ref{ch:Mpool sec:limit to eqm}).}
    \label{fig: BSE convergence}
\end{figure}
\begin{table}\centering
\resizebox{0.9\linewidth}{!}{%
    \begin{tabular}{|l|c|c|c|}
    \hline
     & ${\it BSE}$ & Oceanic Crust & Coreless \\
    \hline
    \ce{SiO2} & 45.97 & 50.36 & 28.8 \\
    \ce{MgO} & 36.66 & 7.61 & 18.7 \\
    \ce{Al2O3} & 4.77 & 15.85 & 2.00 \\
    \ce{FeO}* & 8.24 & 9.56 & 48.7 \\
    \ce{CaO} & 3.78 & 12.24 & 1.8 \\
    \ce{Na2O} & 0.35 & 2.77 & - \\
    \ce{K2O} & 0.04 & 0.13 & - \\
    \ce{TiO2} & 0.19 & 1.48 & - \\
    \hline
    \end{tabular}
    }
    \caption[Literature melt compositions]{Melt compositions (mass fractions expressed as percentages) used for our lava-pool atmosphere evolution model. {\it BSE} (Bulk silicate Earth composition comes from \citet{palme_oneill2003}, oceanic crust from \citet{klein2005crust} and the coreless planet composition from \citet{Elkins-Tanton2008}. *We follow \citet{Kite16} in assuming that the iron oxide is in the form of \ce{Fe2O3} for the coreless composition, otherwise it is in the form of \ce{FeO}.} \label{tab: melt_compositions}

\end{table}
We now investigate the chemical evolution of the lava pool--atmosphere system, using the full model described in \autoref{ch:Mpool sec:model}, for some physically relevant melt compositions. All the compositions used can be found in \autoref{tab: melt_compositions}.

\subsection{Bulk silicate Earth} \label{ch:Mpool sec:BSE}
We start by showing the evolution of a mantle with bulk silicate Earth composition \citep[{\it BSE}, ][]{palme_oneill2003}. This composition is frequently used in the literature as a starting point estimate for lava pool compositions \citep[e.g.,][]{Schaefer_fegley2009,zilinskas22,VapoRock,LavAtmos}.

The evolution of the melt and atmospheric compositions are shown in \autoref{fig: BSE evolution}. We have deliberately highlighted the most significant species. It is useful to compare to the simplified setup in \autoref{fig: simplified_pool}, as the features are essentially the same. Magnesium and iron correspond to the major species of lower and higher volatilities, respectively. Meanwhile, aluminium corresponds to the refractory and sodium (and potassium) to the volatile minor species. The curves of melt and atmospheric concentration are very similar to the simplified model.

\autoref{fig: total_pressures} shows the evolution of the total pressure (blue dotted line) for this model, and models discussed later. This also shows the same features as the simplified model: a monotonic decrease but with bumps as different species dominate the evolution. 

It is important to highlight the behaviour of calcium and silicon in this model. Calcium starts with a low concentration in the melt, but its concentration increases as it is refractory. Eventually, when its concentration is equal to silicon, it decreases, matched with silicon. This is because of a limitation of the chemical model \texttt{MELTS}, that underlies \texttt{LavAtmos}. \texttt{MELTS} models the rock/lava as a mixture of end-member components. In \texttt{MELTS}, calcium can only be present as $\ce{CaSiO3}$, thus the amount of silicon cannot be less than the amount of calcium. This is clearly a major limitation in our model in terms of calculating the actual lava compositions for pools that have undergone significant evolution. As silicon is more volatile, it should be removed and calcium should be in the form of $\ce{CaO}$. Nevertheless, the overall picture of the way that the pool evolves is still reasonable.

\autoref{fig: BSE convergence} demonstrates the convergence of the pool and atmosphere to a steady state composition, where the composition of the atmosphere is the same as the mantle (the lava pool, which the mantle melts into, has a very different composition). More volatile elements, such as potassium, are much quicker to reach the equilibrium concentration, as one might expect. Oxygen is always close to the equilibrium abundance, as all that controls its abundance in the atmosphere is the stoichiometric ratio of the oxides that are vaporised. In other words, at least one oxygen atom is produced per cation (e.g., from \ce{MgO}) and at most two (e.g., from \ce{SiO2}), thus its atmospheric abundance can never vary by more than a factor of 2. 

\subsection{Oceanic crust/`incompatible element' enriched}\label{ch:Mpool sec:oceanic crust}
\begin{figure}
    \centering
    \includegraphics[width=\linewidth]{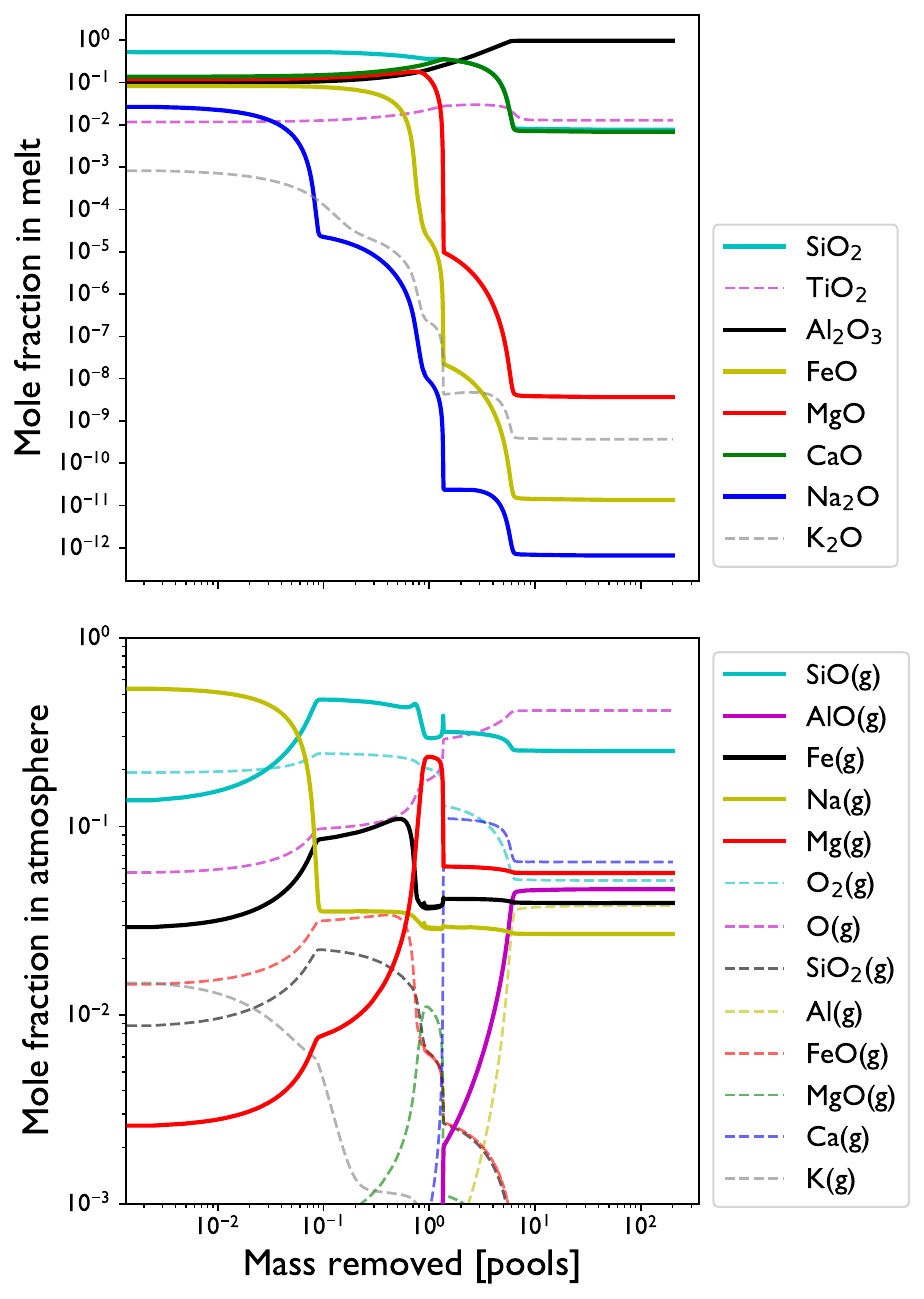}
    \caption[Evolution of the lava pool--atmosphere for melt with oceanic crust compostion]{Evolution of the composition of the melt and atmosphere for a pool initially composed of material with the composition of oceanic crust, and replenished with the same composition at 2600 K. Only vapour species that have a mole fraction above 1\% at some point in the atmosphere's evolution are shown.}
    \label{fig: oceanic evolution}
\end{figure}

By assuming that the melt added to the pool is {\it BSE} in the previous section, we implicitly assumed not only that the mantle has that composition but also that the composition is unchanged as it melts into the lava pool. However, it is possible that the process of melting into the pool causes a compositional change. Under partially molten conditions elements separate into those that dwell preferentially in melt -- so called `incompatible elements' -- and in solid \citep[e.g.][]{Geo:paths+processes}. Unless there are very specific conditions, there should be a partially molten region between the lava pool and the solid deep interior of the planet, where liquid, incompatible element-enriched, material can rise into the pool, as it is typically less dense than the solid. Furthermore, since the planet likely started molten, as crystallisation of the mantle occurred, this process of separation could have happened continually. This is known as fractional crystallisation when considering the solidification of an initial magma ocean \citep{SOLOMATOV-chapter}.

To emulate this potential enhancement in incompatible elements, we consider material with a composition similar to that of the Earth's crust since it is precisely this process of fractional melting which results in the crust's composition \citep[e.g.,][]{klein2005crust}. Specifically, we use an \textit{oceanic crust} composition, which is more appropriate as it is formed directly from magma reaching the surface, as opposed to \textit{continental crust} which has undergone further processing \citep[e.g.,][]{kelemen2016}. One could, in principle, use \texttt{MELTS} (or some other thermodynamic code) to model partial melting to give a composition of material that enters the pool, but the exact conditions in the lava pool are not well enough understood to make this worthwhile. Note that the physical scenario we are trying to reflect is not the mantle being incompatible element-enriched material, but that, in the process of the mantle melting into the lava pool, the material that enters the pool becomes incompatible element-enriched. Of course, using a crust composition also models a situation where the pool is simply formed from melting of the crust. This might be the case early on in a planet's evolution when the planet has not evaporated that much \citep[e.g.,][]{Kite16}.

The evolution of the pool and atmosphere for oceanic crust, at 2600 K, is shown in \autoref{fig: oceanic evolution}. We once again highlight a few atmospheric species to show the general types of evolutions, which are very similar to the {\it BSE} case (\autoref{ch:Mpool sec:BSE}) and the simplified model (\autoref{ch:Mpool sec:simplified}). The spike in the abundance of \ce{SiO} in the atmosphere between 1 and 2 pools worth of mass removed is due to the loss of magnesium relative to calcium and silicon, due to its higher relative volatility. At that point, the abundance of silicon is locked to that of calcium (see \autoref{ch:Mpool sec:BSE}), so its abundance in the melt cannot be depleted as much as magnesium. Thus the silicon to magnsium ratio in the melt and consequently the atmosphere is increased rapidly. One should recall that it is atmospheric fractions that are plotted, not partial pressure, which is decreasing for both species. Since it is due to this locking of the silicon abundance to calcium, the behaviour is likely not physical; however, it is not a numerical error.

The principal difference between the initial pool compositions of crust and {\it BSE} is that crust is much more enriched in aluminium, sodium and to an extent iron. This naturally means the final composition of the atmosphere is more enriched in these. Over the evolution, however, similar species dominate (Na, SiO, Mg) because, until the equilibrium composition is reached, it is the volatility that is more important in determining atmospheric composition.

In both cases, the dominant species in the melt at late points in the evolution is \ce{Al2O3}. Oceanic crust starts with a higher \ce{Al2O3} abundance and thus its progression to the equilibrium state requires a lower amount of mass to be lost. This is best seen by comparing the total pressure curves in \autoref{fig: total_pressures} and seeing that the crust composition reaches a steady state after less mass loss.

\subsection{Coreless planet} \label{ch:Mpool sec:coreless}
\begin{figure}
    \centering
    \includegraphics[width=\linewidth]{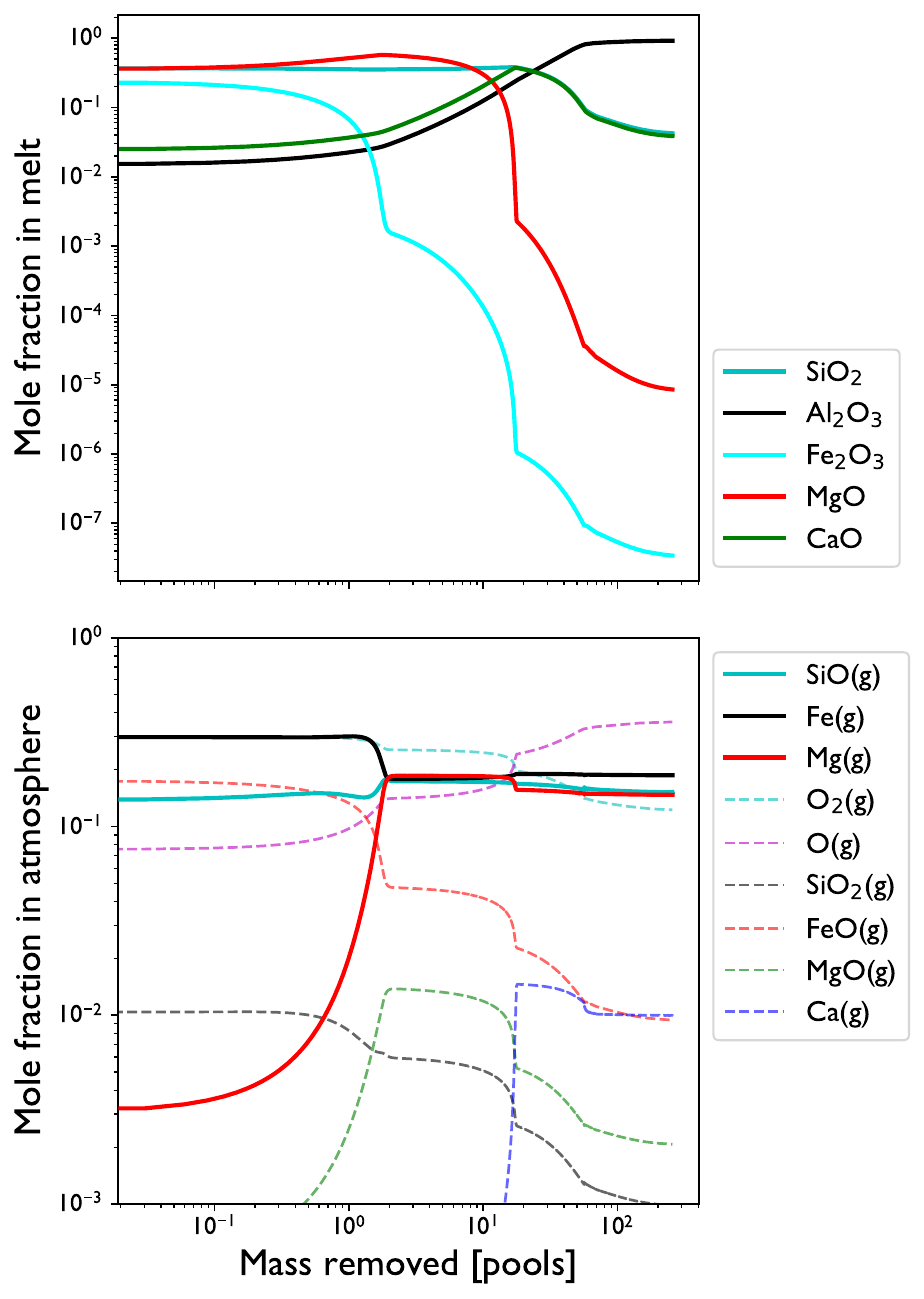}
    \caption[Evolution of the lava pool--atmosphere for a mantle that has not formed a core]{Evolution of lava pool and vapour compositions for a mantle that has not formed a core at a temperature of 2600 K. All melt species are shown. Only vapour species that have a mole fraction above 1\% at some point in the atmosphere's evolution are shown.}
    \label{fig: coreless evolution}
\end{figure}
For our final composition, we consider a planet which has not formed a core. This iron-enriched composition may occur if the planet has insufficient melting for the iron and silicates to separate or if the mantle is very oxidising \citep{Elkins-Tanton2008}. As such a composition is common in an oxidising environment, we follow \citet{Kite16} in assuming that the iron is in the form of iron (III) oxide (\ce{Fe2O3}). This will not make a qualitative difference, although we do return to consider oxidation state in \autoref{ch: Mpool sec: pool_redox_evo}. An enhanced abundance of iron could also occur deep in the mantle due to density stratification \citep[e.g.,][]{Elkins-Tanton2003}, although this would be unlikely to have this exact composition. This might be the material that would melt into the lava pool during the catastrophically evaporating stage when the upper mantle has already been lost. 

The evolution, again at 2600~K, is shown in \autoref{fig: coreless evolution}. In this case, the trace alkali metals and titanium are not included. Evolution is similar in form to the previous cases. Because of the enhanced iron abundance, it is the dominant gas species (other than oxygen) for a large part of the evolution\footnote{For a real planet, at early stages, sodium would be dominant as it has been for previous cases if it had been included.}. Because of low \ce{Al2O3} content, it takes longer for equilibrium to be reached than the previous cases, again seen best by looking at the total pressure evolution in \autoref{fig: total_pressures}.

\subsection{Temperature variation}
The evolutions shown in the previous sections have all been for 2600~K. In \autoref{fig: total_pressures}, we also show some evolutions of total pressure for lower temperatures to demonstrate some associated trends. There are two main trends. Firstly, lower temperatures correspond to lower partial pressures due to less vapour being produced. Secondly, for those compositions with sodium ({\it BSE} and oceanic crust), the pools undergo their early evolution at lower temperatures with a lower amount of mass removed. This is because, in the temperature range considered, sodium's volatility has a less steep dependence on temperature than other species (see plots of composition with temperature in, for instance, \citealt{LavAtmos}). This means at lower temperatures it is relatively more volatile and thus it is removed more readily. For the rest of the evolution, similar effects due to the temperature dependence of different species' partial pressures can occur but are much more subtle. It is important to recall that the $x$-axis is not time but mass removed. Higher-temperature planets lose mass more quickly; thus, were the $x$-axis time, the higher-temperature planets would evolve much more quickly (see \autoref{ch:Mpool sec: likelihood of evolved}.)

An interesting effect that occurs at low temperatures is that sometimes the composition of the melt can evolve to a point when the equilibrium assemblage, as calculated by \texttt{MELTS}, is not pure liquid, but contains some solid. We mark when this has occurred by a cross in \autoref{fig: total_pressures}. In principle, the model can continue evolving by using only the liquid part of the assemblage (as occurs for the coreless 2000~K case.) In practice, the model often fails to converge because of the composition change as solid forms (as occurs for the {\it BSE} and oceanic crust cases), and since precisely what should occur at the point of partial solidification is not clear, we have not attempted to extend beyond the point when solid forms. As noted in \autoref{ch:Mpool sec:BSE}, the exact melt compositions that are calculated should not be used to infer those of real lava pools, due to limitations of \texttt{MELTS}. However, it could be that generic partial solidification of the lava pool, due to its compositional evolution, is a real effect that should be taken into account. The exact consequences would require modelling of the mixing/settling or even floating of the solid produced. The solid material produced in all our models is denser than the liquid\footnote{The solids produced are melilite for {\it BSE} and oceanic crust at 2000 K, corundum for {\it BSE} at 2300 K, melilite for oceanic crust at 2300 K, and olivine for coreless at 2000 K.}, so should sink, but this is not necessarily true for all compositions. 

\subsection{Potential oxidation evolution of the lava pool}\label{ch: Mpool sec: pool_redox_evo}
\begin{figure}
    \centering
    \includegraphics[width=\linewidth]{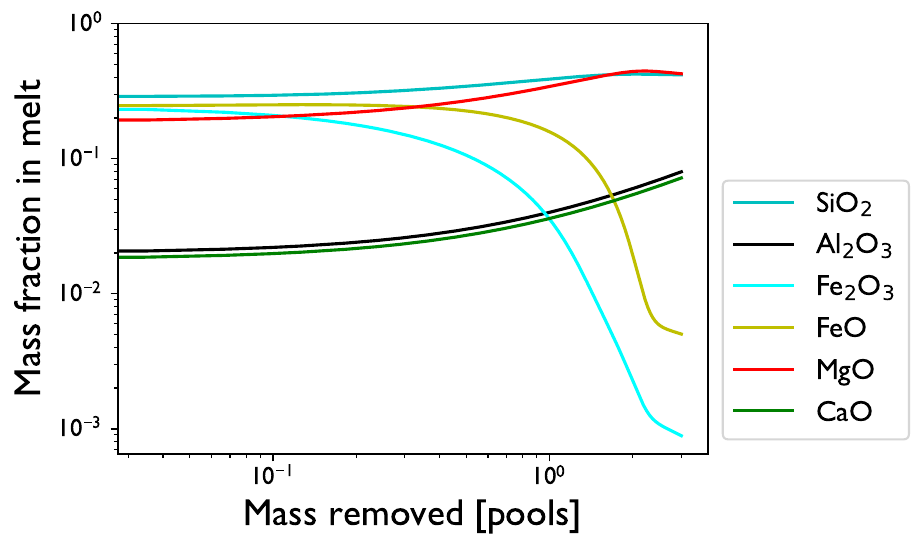}
    \caption[Comparison of \ce{FeO} and \ce{Fe2O3} evaporation]{Evolution of the composition in the melt and atmosphere for a pool fed by a mantle with the same composition as the coreless planet, but with iron partitioned in an equal mass of \ce{FeO} and \ce{Fe2O3}. Evolution was run for 2600~K, for three pools worth of mass removed. Note that the melt composition is given as a mass fraction.}
    \label{fig: 2oxides}
\end{figure}
Here we briefly consider the different behaviour if one assumes that iron in the mantle is in the form of both ferrous iron (II) oxide (\ce{FeO}) and ferric iron (III) oxide (\ce{Fe2O3}). The \texttt{MELTS}/\texttt{LavAtmos} framework we use to calculate partial pressures of vapours does not include oxidation reactions to convert these oxidation states of iron between each other, so any evolution involving both forms will not be quantitatively true, but may still hold some interesting insight. To investigate this we evolved a pool and atmosphere model where we took the composition of the coreless planet used in \autoref{ch:Mpool sec:coreless}, but separated the iron into equal mass portions of \ce{FeO} and \ce{Fe2O3}. This is shown in \autoref{fig: 2oxides}.

\autoref{fig: 2oxides} shows that iron is more volatile in the form of \ce{Fe2O3}, as the mass fraction of this in the melt decreases more rapidly than that of \ce{FeO}. This difference in volatility may influence the evolution of the oxidation state of the lava pool. As already stated, iron (II) to iron (III) oxidation reactions are not included, so it is possible that the actual evolution of the lava pool occurs differently. However, it does highlight a need for consideration of how fractional vaporisation affects redox evolution. This is of particular significance as there is interest in inferring mantle oxidation states from atmospheric observations \citep[e.g.,][]{VapoRock}.

\section{Likelihood of evolved lava pools}\label{ch:Mpool sec: likelihood of evolved}

In the previous sections, we showed that fractional vaporisation significantly alters the composition of lava pools and their atmospheres. The most evolved state is one where the atmospheric composition is the same as the material that is melted into the lava pool (either that of the mantle or incompatible element enriched melt from the mantle). At intermediate points in the evolution, the atmospheric composition is defined by a balance between the volatility of species and their abundance in the melt. We have deliberately presented the evolution results as a function of the amount of mass removed from the pool, in order to show the general results independent of the mass loss process. Generally, it only takes $\sim10 - 100$ pools worth of material to be removed before the final equilibrium composition is reached.  In this section, we attempt to gauge for which planetary conditions (temperature, size and age) the pools are likely to be in the evolved equilibrium state, i.e., when have they lost $\sim10 - 100$ pools. The uncertainties in the models do not allow more precise evolutionary stages to be determined.

In order to work out for what conditions planets' lava pools are in the evolved stage, one simply needs to know the mass contained in the lava pool and how fast it is removed. Unfortunately, neither of these quantities are known precisely, but, in the following subsections, we will make estimates of them.

\subsection{Lava pool depth}\label{sec: pool depth}
An important ingredient to working out the likelihood of pools becoming evolved is their depth, and thus mass. The dayside surfaces of lava planets are above the melting point of rocks, but their interiors are not \citep{CURRY2024}. The base of the lava pool is best defined as the point where the rock reaches the critical melt fraction, where a sharp transition occurs between solid and melt-like behaviour \citep[e.g.,][]{SOLOMATOV-chapter}. To find this point, one really needs to know the full (3D) temperature structure of the planet, which is difficult. 

\citet{nguyen2020} assume a vertically constant temperature structure below the lava pool's surface and produce pools of $\sim150$ km. \citet{boukare2022}, meanwhile, assume that energy is flowing out from the deep interior and so construct convective and conductive profiles to argue for deep pools (100s km). However, when one considers heat transport in the planet in more than one dimension, there is no reason that at all angles heat should be flowing outwards. Indeed the lava pool will likely have higher entropy than the interior of the planet \citep{vanSummeren2011,Kite16}, especially once the interior of the planet has cooled \citep{CURRY2024}. If one assumes conduction into the planet, with some internal temperature, one finds shallower pools (10s km - see \citealt{Kite16}, Appendix B, and \citealt{CURRY2024}, section 4.1).

The estimate of lava pool depth is further reduced by the effect of circulation in the lava pool, as argued by \citet{Kite16}. The pool should have a steep thermal boundary layer at the surface, given by
\begin{equation}
    \delta_T = \left(\frac{k}{\rho C_P} \frac{4 \Omega \sin(\theta_p/2)\tan(\theta_p/2) R^2\theta_p}{(\Delta\rho/\rho) g} \right)^\frac{1}{3} \label{eq: pool BL}
\end{equation}
\citep[adapted from][Eq. 8]{Kite16}, where $k$, $C_P$ and $\rho$ are the conductivity, specific heat capacity and density of the pool material; $R$, $\Omega$ and $g$ are the planet radius, period and surface gravity; $\theta_p$ is the angular radius of the pool and $\Delta\rho$ is the density change from pool centre to edge (see also \citealt{CURRY2024}, section 4.1). \citet{Kite16} argue that the total depth of the pool ($\delta_p$) should be no more than $10\times$ this depth. This is because the lower temperature of the boundary layer should be close to the solidification temperature (or more precisely the point where the melt fraction is low enough that the material behaves like a solid). For the rest of this section, we will use this ($\delta_p = 10\delta_T$) as our estimate of the pool depth.

To evaluate \autoref{eq: pool BL}, we assume that the surface temperature is given by \citep[e.g.,][]{leger2009}
\begin{equation}
    T(\theta) = T_{ss}\cos^\frac{1}{4}{\theta} \; , \label{eq: T(theta)}
\end{equation}
where $T_{ss}$ is the `substellar temperature' -- the maximum temperature of a bare tidally locked planet (i.e., that closest to the star), the surface of which acts as a local black body.
\begin{equation}
    T_{ss}^4 \equiv T_*^4 \frac{R_*^2}{a_\text{orbit}^2} \; , \label{eq: Tss}
\end{equation}
where $T_*$ and $R_*$ are the stellar surface temperature and radius and $a_\text{orbit}$ is the semi-major axis of the planet's orbit.
$\theta_p$~is calculated by finding at what angle from the substellar point that material solidifies (which occurs at a temperature $T_\text{edge}$), so
\begin{equation}
    \theta_p = \arccos(\frac{T_\text{edge}^4}{T_{ss}^4}) \; . \label{ch:Mpool eq: pool_size}
\end{equation}
Following \citet{Kite16}, we define $T_\text{edge}$ to be the temperature when material switches from behaving as a liquid to a solid (it will technically still be partially molten) and use their value of $T_\text{edge}=1673$~K.

Using our assertion that the pool depth is small relative to the planet's radius, and assuming a constant depth across the pool, the pool mass is
\begin{equation}
    m_\text{pool} = 2\pi R^2 (1 - \cos{\theta_p}) \delta_p \, \rho \; .
\end{equation}
The pool mass scales linearly with the pool depth, because of the assumption of its small size relative to the planet. Increasing $\theta_p$ increases the azimuthal size of the pool and thus its mass and the spherical geometry is taken into account by the cosine factor. Since $\theta_p$ is set by the stellar irradiation (\autoref{ch:Mpool eq: pool_size}), as the planets radius increases the mass of the pool also scales with the surface area of the planet, hence the $R^2$ factor.
\subsection{Amount of mass removed from planets}\label{ch:Mpool sec:mass_removed_from_planets}
\begin{figure}
    \centering
    \includegraphics[width=\linewidth]{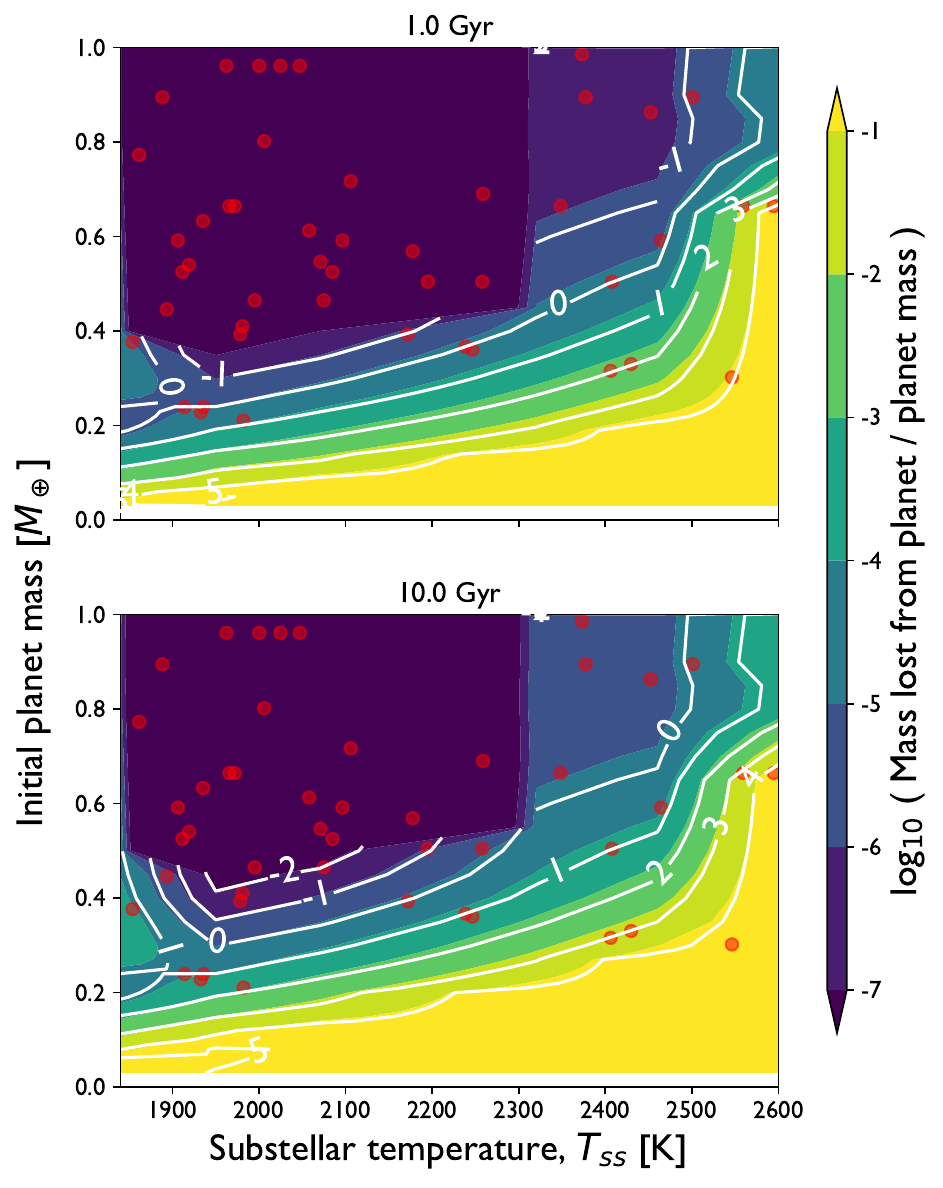}
    \caption[Mass removed from planets compared to pool mass]{The amount of mass removed from planets with different initial masses and substellar temperatures at different ages (the top panel corresponds to 1~Gyr and the bottom to 10~Gyrs). Planets lose mass according to the models of \citet{Booth_disint22}. Colours plot mass lost / initial planet mass while white contours plot log$_{10}$( mass lost / pool mass) at the specified age. See the text for a discussion of the pool mass. Red circles show the estimated {\it current} mass of planets in the exoplanet archive. (These are generally large planets, that will not change significantly in mass over their lifetimes). The box-like shape of the colour contours in the top left is an artefact of the finite grid of temperatures that the mass loss models were run for.}
    \label{fig: m_lost/mp Booth}
\end{figure}
We first consider the amount of material fully removed from the planet. To investigate this, we ran the interior structure model described in \citet{CURRY2024}, coupled to the mass loss model of \cite{Booth_disint22}, over a range of planet temperatures and initial masses. Because we were not interested in the thermal evolution, and \citet{CURRY2024} showed that molten planets, even under strong irradiation, tend to crystallise within 10,000 years, we set the thermal structure to be adiabatic, with an outer mantle temperature of 1200~K. In \autoref{fig: m_lost/mp Booth}, we show the amount of material lost by planets as a function of their initial mass and substellar temperature, $T_{ss}$, for different time snapshots. Colours show the mass lost in terms of the total planet mass at that time; white contours show the mass removed divided by the mass of the lava pool.

This figure highlights that, to get a significant evolution of the lava pool, which just requires 10-100 pools worth of material to be removed, only a small fraction of the entire planet needs to be removed. For instance, 100 pools worth of material is removed when only around 0.1\% of the total mass is lost. Consequently, relatively large planets can have significant pool evolution: at 2600 K, a 0.7$M_\oplus$ planet can lose 100 pools worth by 1 Gyr. There is, of course, great uncertainty in the actual pool depth, but it is certainly much shallower than the size of the planet, so the general conclusion is robust. 

The catastrophically evaporating planets have temperatures between 2000-2300~K, and are expected to have masses $<0.1M_\oplus$ \citep{Perez-Becker13,CURRY2024}, so all sit in an evolved regime. This means we should expect the material in their outflows to reflect the composition of the material that is melted into their lava pools. This potentially means that measuring the tail composition is a direct measure of mantle material, as opposed to one modified by volatility. As discussed in \autoref{ch:Mpool sec:oceanic crust}, however, this may be modified by the process of melting.

\subsection{Mass moved from the dayside to the nightside}
\begin{figure}
    \centering
    \includegraphics[width=\linewidth]{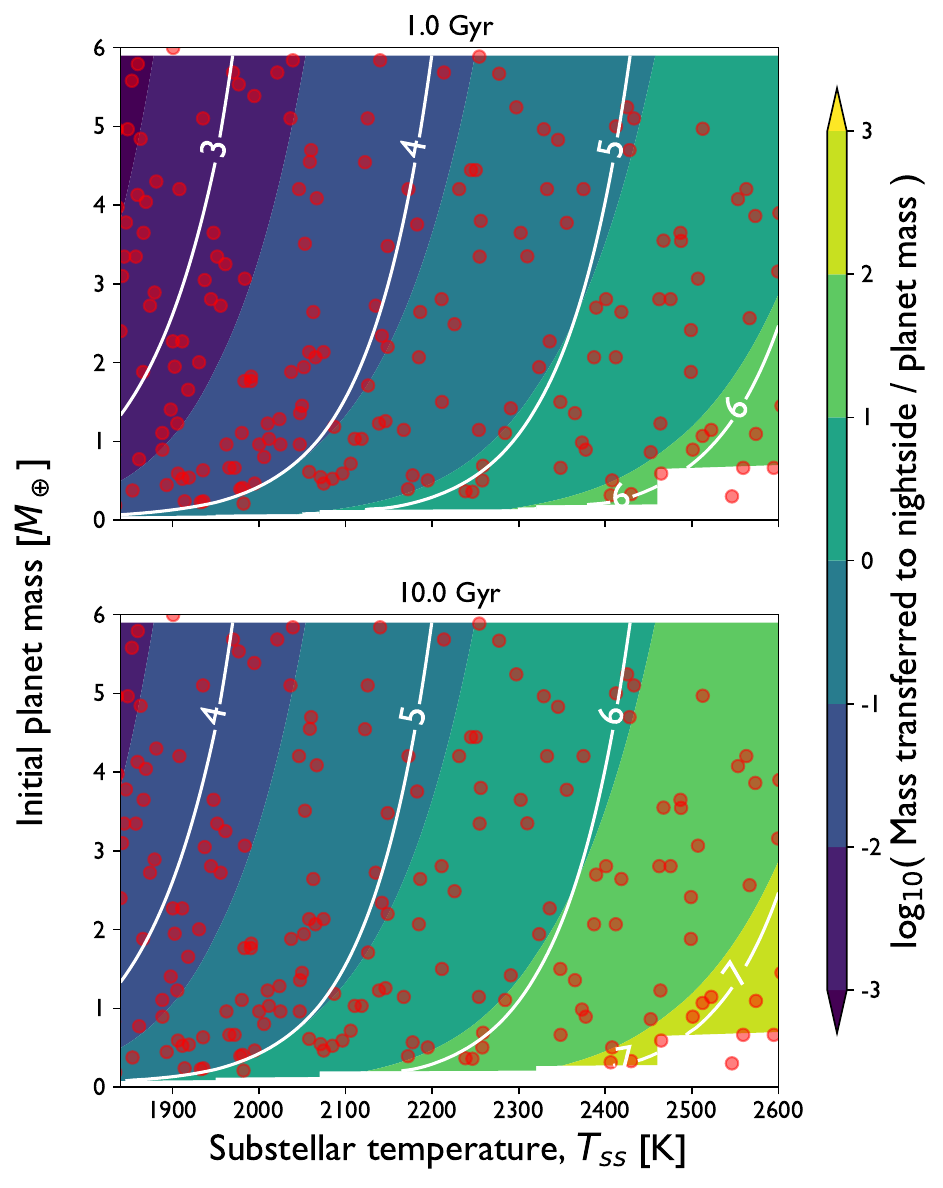}
    \caption[Mass moved to the nightside compared to pool mass for \ce{SiO}]{The amount of mass removed from lava pools via day-to-nightside winds for planets of different initial masses and substellar temperatures and at different ages. The composition of the material removed in the day-to-nightside wind is assumed to be \ce{SiO}. The planets' total masses also decrease due to mass loss according to the models of \citet{Booth_disint22}. The colours show the amount of mass removed / the initial mass of the planet, while the white contours show log$_{10}$( amount of mass removed / pool mass ) at the specified age. See the text for a discussion of the pool mass. Red circles show planets in the exoplanet archive.}
    \label{fig: m_lost/mpool day-to-night Booth}
\end{figure}
\begin{figure}
    \centering
    \includegraphics[width=\linewidth]{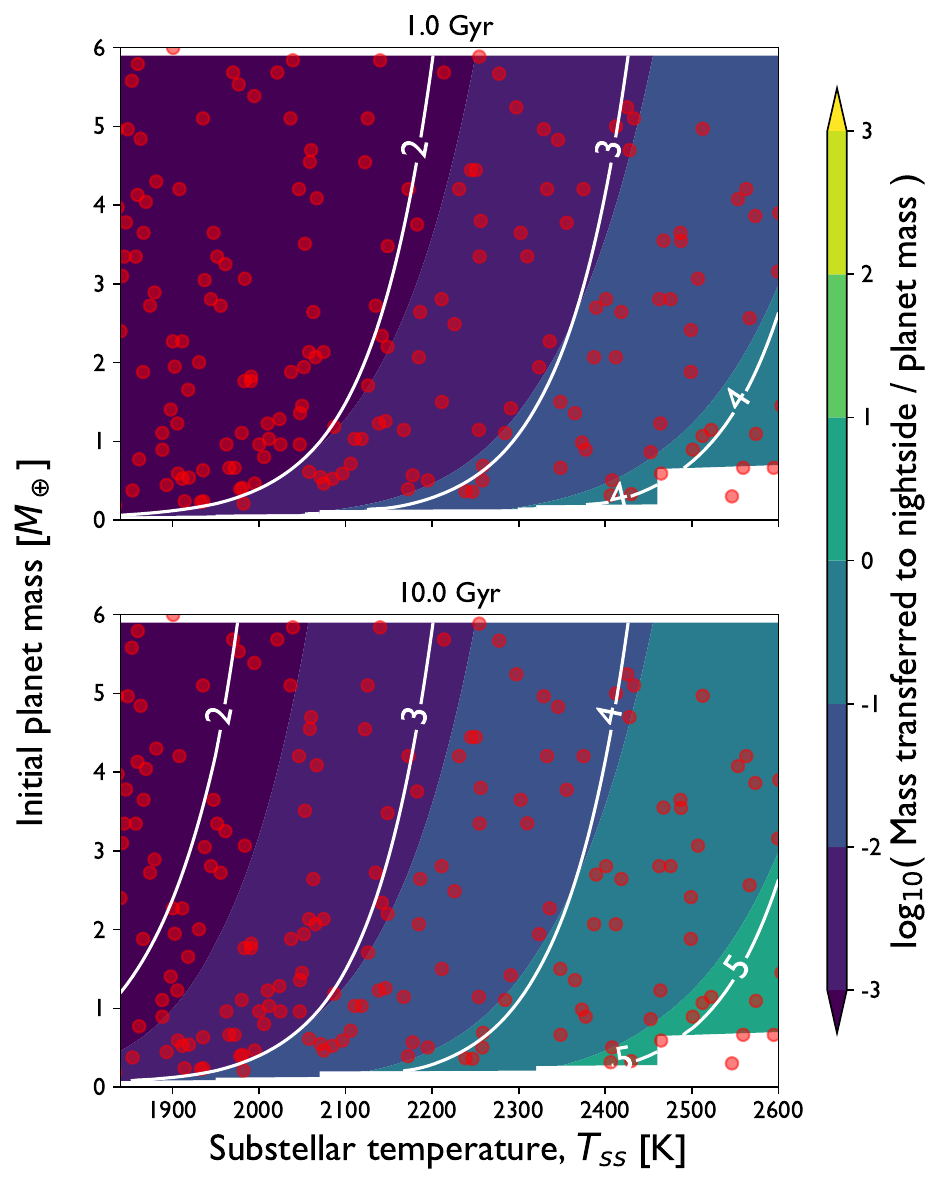}
    \caption[Mass moved to the nightside compared to pool mass for \ce{Al2O3}]{Same as \autoref{fig: m_lost/mpool day-to-night Booth} but instead the wind composition is assumed to be \ce{Al2O3}. The colour scale is also the same as \autoref{fig: m_lost/mpool day-to-night Booth}.}
    \label{fig: m_lost/mpool day-to-night Booth Al}
\end{figure}
For the composition of the lava pool to change due to evaporation, one does not actually need the mass removed from the pool to leave the planet entirely. Tidally locked lava-pool planets are believed to have strong day-to-nightside winds \citep{castan-menou11}, and this day-to-night transport could cause the compositional evolution instead. 

The amount of material that passes out across the edge of the pool (which is approximately that which passes across the terminator) may be estimated as \citep[see][section 5.2]{Kang2021}
\begin{equation}
    \dot{m}_\text{term} \sim 2\pi \bar{P} \sin{\theta_p} \frac{R}{g} \sqrt{\frac{k_B\bar{T}}{\mu}} \; , \label{ch:Mpool eq:day-to-night mdot}
\end{equation}
where $\bar{P}$ and $\bar{T}$ are the average atmospheric pressure and temperature over the pool, $\mu$ is the mean molecular weight of the atmosphere and other quantities are as defined in \autoref{sec: pool depth}.

$\bar{P}$ may be estimated as the equilibrium pressure of an atmosphere above the lava pool at a temperature $\bar{T}$. This is precisely what we were working out with \texttt{LavAtmos} in \autoref{ch:Mpool sec: calc atm comp}, although often the functional form
\begin{equation}
    P_\text{eq} = A \exp(-B/T) \label{ch:Mpool eq:Peq formula}
\end{equation}
is used \citep{castan-menou11,Kang2021}, where $A$ and $B$ are composition dependent constants. It is interesting to note that the mass and size of the planet only comes into \autoref{ch:Mpool eq:day-to-night mdot} through $R/g$, which is approximately constant (since $g \appropto R^3/R^2 = R$). This means that for larger planets, day-to-night transport does not decrease that much, in contrast to mass loss to space, so it has a larger impact for those planets\footnote{An additional consequence of this is that mass loss from the nightside is more important for larger planets, and from the dayside for smaller planets \citep{Kang2021}.}.

In order to address the problem of how much mass is transported horizontally in full, one should really consider the compositional evolution. This would also require choosing the composition so, for simplicity, we will just stick to one species and use \autoref{ch:Mpool eq:Peq formula} to give the pressure of the atmosphere above the lava pool. We show results for two compositions, \ce{SiO} and \ce{Al2O3}. \ce{SiO} is the dominant outgassed species (other than oxygen) for a large period of lava pool evolution for multiple compositions, as demonstrated in Figures \ref{fig: BSE evolution} and \ref{fig: oceanic evolution}. \ce{Al2O3}, meanwhile, is the pool composition at late times, and thus assuming a pool--atmosphere composed of it models mass loss from a highly evolved lava pool. It also gives a lower estimate of the amount of material that can be removed since it is a more refractory species. We used values of $A$ and $B$ for \autoref{ch:Mpool eq:Peq formula} from \citet{Kang2021}, which implicitly assumes melt with bulk earth composition. 

To find $\bar{T}$ for a given substellar temperature, we assumed that the surface temperature is given by \autoref{eq: T(theta)}, which yields
\begin{equation}
    \bar{T} = \frac{4}{5} \, T_{ss}  \left(\frac{1 - \cos^\frac{5}{4}{\theta_p}}{1 - \cos{\theta_p}}\right) \; . \label{ch:Mpool eq:avg_T_pool}
\end{equation}
$\theta_p$ is again found using \autoref{ch:Mpool eq: pool_size}.

\autoref{fig: m_lost/mpool day-to-night Booth} and \autoref{fig: m_lost/mpool day-to-night Booth Al} show the mass removed from the lava pool horizontally under these assumptions, at different planet initial masses and substellar temperatures, for \ce{SiO} and \ce{Al2O3} winds respectively. One can see that far more material is removed horizontally than is lost from the planet entirely in \autoref{fig: m_lost/mp Booth}. This is true even if the very refractory composition of \ce{Al2O3} is assumed. The consequence of this is that far more planets are likely to be in a regime where their lava pools are highly evolved. Indeed, even planets of several Earth masses can have highly evolved lava pools, meaning it is likely that this evolution needs to be taken into account when interpreting observations of the atmospheres of {\it all} lava planets, as we will discuss in \autoref{sec: atm obs}. This stretching of the evolved regime to high-mass planets is essentially a consequence of the fact that day-to-nightside transport does not depend highly on planet mass, as discussed above.

Not only can planets' lava pools become highly evolved, but \autoref{fig: m_lost/mpool day-to-night Booth} suggests that planets can have their entire masses transported to the nightside multiple times over. Even for refractory \ce{Al2O3}, this is the case for high temperatures. For low-mass or high-temperature planets, \citet{kang2023-TPW} show that planets are likely to undergo true polar wander due to the mass redistribution, meaning the process of pool evolution will be periodically reset. However, for planets $\gtrsim 1 M_\oplus$, this is unlikely due to a longer timescale of mass build-up compared to mantle relaxation. In these cases,  mass must somehow circulate back through the planet. Consideration of the consequences of this for the lava pool evolution is beyond the scope of this work. 

\section{Discussion}\label{ch:Mpool sec:discuss}

\subsection{Mass of atmosphere compared to pool}\label{ch:Mpool sec:atm mass}
\begin{figure}
    \centering
    \includegraphics[width=0.9\linewidth]{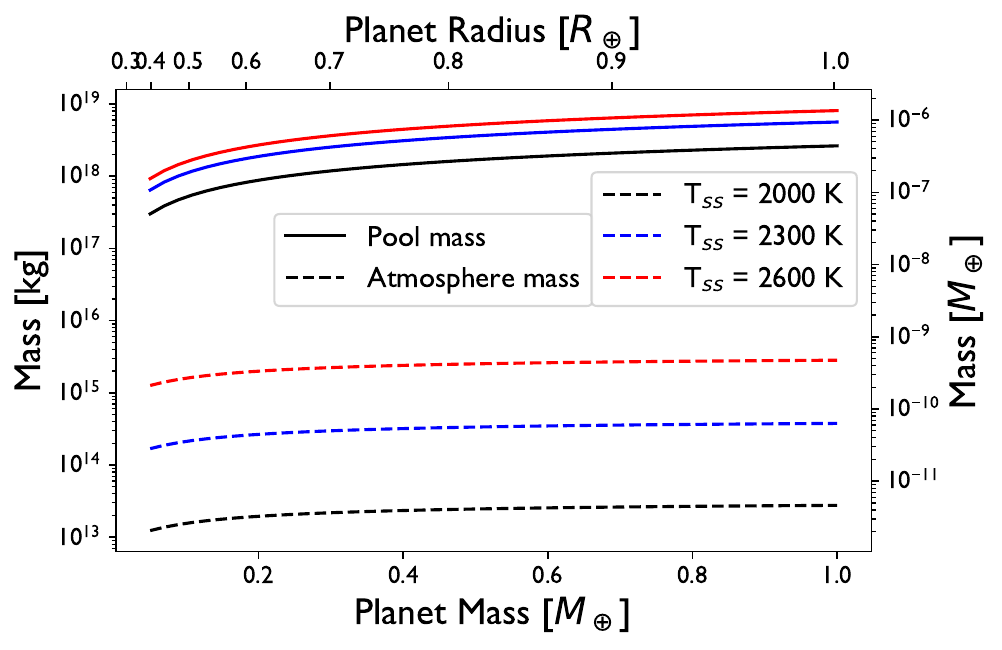}
    \caption[Comparison of pool and atmospheric masses]{Comparison of pool and atmospheric masses for planets with lava pools with oceanic crust composition for planets with different substellar temperatures, $T_{ss}$ and planet masses. See text for further calculation details.}
    \label{fig: m_pool_vs_m_atm}
\end{figure}
In \autoref{ch:Mpool sec: calc atm comp}, we asserted that the atmospheres generated have a smaller mass than the lava pools they are evaporated from. This can only be verified once the atmospheres are generated, and we do so here. The mass of the atmosphere can be estimated as
\begin{equation}
    m_\text{atm} = \frac{P_\text{tot}}{g} A_{atm}
\end{equation}
where $P_\text{tot}$ is the total pressure of the atmosphere, $g$ is, as ever, the gravitational acceleration, and $A_{atm}$ is the area on the surface where there is an atmosphere. This will be approximately the area of the pool. To verify that, even in relatively extreme cases, the pool mass is larger, we ran calculations of atmospheric and pool masses, where we deliberately maximised the mass of the atmosphere and minimised the mass of the pool. To maximise the mass of the atmosphere, we used oceanic crust composition melt as it gives the most massive atmosphere of the compositions we consider (see \autoref{fig: total_pressures}.) We also assume that the pool depth is equal to the size of the thermal boundary layer in a convecting pool ($\delta_T$, see \autoref{sec: pool depth}), which is very much a lower limit to the pool mass. The planet radius is calculated by assuming the mass-radius formula in \citet{Fortney07}, which allows the calculation of $g$. As in the previous section, the substellar temperature and pool size are calculated using Equations \ref{ch:Mpool eq:avg_T_pool} and \ref{ch:Mpool eq: pool_size}. Pool and atmospheric masses for different planet masses are plotted in \autoref{fig: m_pool_vs_m_atm} and one can see that it is indeed the case that the pool masses are orders of magnitude larger.

\subsection{Changing depth of the lava pool}\label{ch:Mpool sec:non const mass}
For the calculations in this work, we have assumed that the mass of the lava pool stays the same. In reality, this should not be the case. There are three factors which can change the pool's depth. Firstly, the total mass of the planet will affect the mass of the pool; in the catastrophically evaporating cases, the planet's mass changes dramatically. Secondly, the pool's depth will depend upon its composition. One can see this by inspecting \autoref{eq: pool BL}, which depends on several material properties. Thus as the pool's composition evolves so will its mass. Finally, the atmosphere may alter the depth by changing the surface temperature.

The first point is unlikely to be very important. Firstly, \citet{CURRY2024} showed that the pool depth evolution, based purely on changing mass, is relatively weak (see their Figure 9), especially if pools have efficient circulation, as we described in \autoref{sec: pool depth}. Secondly, as discussed in \autoref{ch:Mpool sec: likelihood of evolved}, because the pools are small relative to the size of planets, the pool evolution is much faster than the mass evolution of the planet, thus one can effectively consider the pool evolution to occur at constant planet mass. The other two points will potentially have an effect. In this work, we have explicitly shown that the composition of the pool changes. Thus the thermal properties of the pool will change and so will its depth. Furthermore, it is believed that lava pool atmospheres can affect their surface temperatures, particularly through optical depth change due to dust production \citep{Booth_disint22,Bromely-Chiang23}. This variability is believed to be fast ($\sim 1$ day) compared to lava pool circulation ($\sim 1$ year, \citealt{Kite16}), so it is unlikely to directly affect the pool's global properties. However, there may be some longer-term trends in the average surface temperature due to the changing atmospheric composition. 

The final state of the pool, however, will be unchanged by any depth alterations during the evolution. The steady state pool composition, by definition, has a constant pool and atmosphere composition. It thus also has a constant depth, since this is determined by the pool and atmosphere properties. If one believes the inferences in \autoref{ch:Mpool sec: likelihood of evolved}, that the steady state is quickly reached by most planets, then the pool depth changing will have no effect on the observed state of those planets. The changing pool depth might affect the timing of reaching the steady state, but it is unlikely to be a large change. This is because the physical properties of pool components will not change drastically, even as composition or temperature changes. There is also a limit to how much the relatively thin atmosphere can change the surface temperature. 

In conclusion, while it is true that when modelling the pool--atmosphere system's evolution in full one should include the evolution of the pool's depth, it will neither change the final state of the pool nor the basic picture of the evolution. 

\subsection{Inclusion of volatile species}\label{sec: volatiles}
Our model has throughout assumed that the volatile elements that make up traditional planetary atmospheres (i.e., C, N, S) are not present. This makes sense as they may simply not condense sufficiently in a planet formed close to the star or can be removed quickly by atmospheric escape \citep{Schaefer_fegley2009}. However, it has been proposed that if the volatiles are incorporated into the solid body (for instance, if the planet formed further out), the atmosphere can be continually resupplied through a magma ocean. This is a possible explanation for the apparent detection of a volatile-rich atmosphere on the ultra-hot super-Earth 55 Cancri e \citep{55cnce-nat24}. As a reasonably high mass planet \citep[$\sim 8 M_\oplus$,][]{Bourrier2018}, volatile retention may be more likely on such a planet than on lower mass equivalents. 

The presence of a volatile-rich atmosphere in itself would not prevent the evaporation of rock-forming elements from the lava pool. However, redistribution of energy to the nightside by the atmosphere would decrease the dayside surface temperature, thus decreasing the rate of lava pool evaporation. If a purely dayside lava pool does exist (it is unlikely that day-night temperature contrasts will be entirely wiped out), then it is very feasible that the lava pool evolution will still proceed as we have described, just within, and modified by, a thicker atmosphere. This can occur in exactly the same manner of escape to space or day-to-night transport as we have discussed in the rest of the paper. Light volatile species that have higher velocities may well aid the evolution by dragging the refractories. Further study is required to understand the interaction between traditional volatiles and the evaporated rock-forming elements, as well as the parameter space for which volatile retention is relevant.

\subsection{Buoyant evaporative residue}
\begin{figure}
    \centering
    \includegraphics[width=\linewidth]{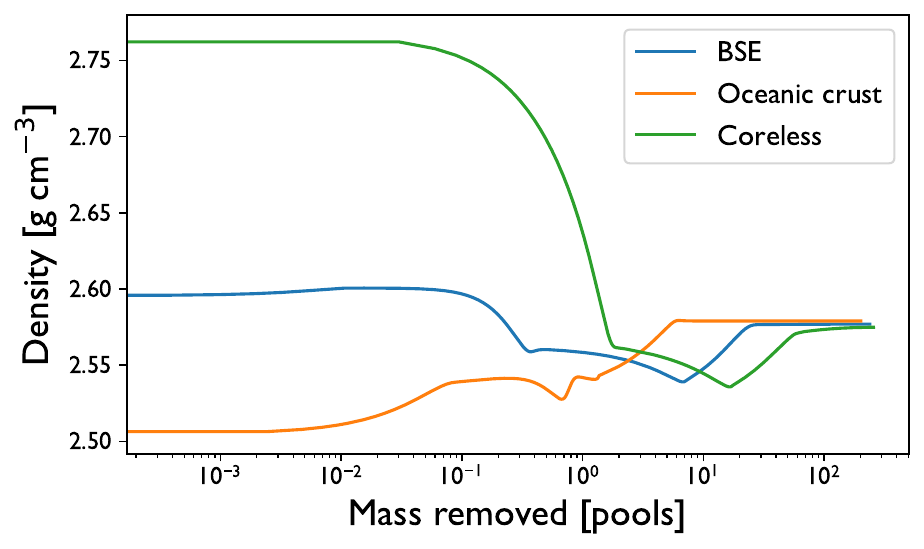}
    \caption[Evolution of the density of lava pool residue]{Density of lava pool residue as material is removed from pools of different initial compositions. Density is calculated using \texttt{MELTS}. All runs are at a temperature of 2600 K.}
    \label{fig: lava pool density}
\end{figure}
\citet{Kite16} investigated another feature of lava pool compositional evolution which can change the end-state of pool evolution. As the pool composition changes its density changes. If the density of the new composition is greater than that of the material melted into the pool and only the upper layer of the pool is altered, then this material will sink. This means that the upper layer of the pool, the part that affects the atmosphere, can only reach this composition where the density is equal to that of fresh melt. On the other hand, if the material is less dense, it will remain at the surface and continue to evolve. If the circulation in the pool is quick, this is the same scenario as we have studied in earlier sections. If circulation is slow, instead of the whole pool reaching a steady state, it is just the upper layers of the pool. This may mean that the assumption that the melt reservoir has a much higher mass than the atmosphere, justified in \autoref{ch:Mpool sec:atm mass}, breaks down.

In \autoref{fig: lava pool density}, we plot the evolution of the density of lava for each of the initial pool compositions we considered in \autoref{ch: Mpool sec: pool_results}. Each behaves rather differently. To decipher the origins of the density changes in detail, one can cross-refernce them with the composition plots (Figures \ref{fig: BSE evolution}, \ref{fig: oceanic evolution} and \ref{fig: coreless evolution}). Notable features are as follows. The oceanic crust density initially increases greatly due to the loss of volatile alkali metals, which are in general lighter. This also occurs for {\it BSE} but to a much lower extent as the proportion of alkali metals is lower, but not for the coreless composition as trace alkali metals are not included. There is then a decrease in density, which is driven by the loss of iron. This is the largest for the coreless composition as it has the largest iron content. In the final stages before reaching a steady state, all three compositions have an increase in density as denser \ce{CaSiO3} and \ce{Al2O3} become dominant. 

In general, this agrees with the results in \citet{Kite16}, although with some quantitative differences. One would not expect them to be precisely the same for a few reasons. For example, we use a different equation of state: we use the inbuilt \texttt{MELTS} equation of state \citep{MELTS1}, whereas \citet{Kite16} uses \citet{Ghiorso+Kress2004}. Furthermore, as discussed above, our framework cannot produce \ce{CaO}, whereas theirs can. Finally, they do not include melt replenishment and the reaching of the steady state. All this being said, the only significant difference between our results is that, for their simulations, {\it BSE} composition increases to above the initial density in the latter stages, whereas ours does not. This means that they predict a point when the composition stops evolving due to the surface material sinking for {\it BSE} while our model would not. This further highlights the need for careful modelling of these evolved pool compositions. We agree, however, that oceanic crust is always denser than its original composition and coreless is always less dense. This may mean that if the composition of the pool is similar to oceanic crust, its composition can never evolve as the surface remains the same as the initial melt. While the fractionated material would sink in the pool, it is not denser than solid material, so it may well get remixed deeper in the pool, meaning evolution would eventually occur. One would need to study the dynamics of the pool in greater detail to ascertain the precise evolution, which is beyond the scope of this work.

\subsection{Connection to the composition of the dusty tails of evaporating rocky planets}
Recent work by \citet{CamposEstrada23} modelling the dust tails of the catastrophically evaporating planets Kepler 1520b (KIC 1255b) and K2-22b suggest that the best-fit composition for the dust is a magnesium iron silicate. Since the mantle is rich in such materials, this agrees with our conclusion that the catastrophically evaporating planets' outflows should match that of the material melted into the lava pool. They disfavour corundum, which had been proposed in earlier works \citep{vanLieshout14,vanLieshout16}, due to a combination of the very high particle sizes and mass loss rates required for it to fit the data and physical arguments based on the fact that Al is a minor constituent of planets. However, if the material melted into the pool were biased towards less-compatible elements, as we argued might be the case in \autoref{ch:Mpool sec:oceanic crust}, then corundum might be a more possible species, although silicates should still be more common (see the crust composition in \autoref{tab: melt_compositions}). As \citet{CamposEstrada23} point out, it should be possible to rule in or out corundum using {\it JWST} spectroscopy and a program to observe K2-22b using MIRI has already been approved \citep{k2-22b_jwst}.

Both transit modelling efforts \citep{CamposEstrada23} and theoretical models of the variability of the tails \citep{Booth_disint22,Bromely-Chiang23} support the specific inclusion of iron in the silicate dust. There are at least three ways that our model can enhance iron. Firstly, it could simply be that the mantle is richer in iron, for instance, because the planet did not form a core (\autoref{ch:Mpool sec:coreless}), due to its low mass. Secondly, the deeper mantle that the catastrophically evaporating planets have evaporated to may be enriched in iron due to compositional stratification (\autoref{ch:Mpool sec:coreless}). Thirdly, fractional melting (\autoref{ch:Mpool sec:oceanic crust}) would also enhance the iron in the outflows. Without more details on the dust, gas, or pool composition, one cannot distinguish between these scenarios. 

The first two scenarios may, in principle, be distinguishable by measuring the atmospheres of the planets or outflowing gas, particularly the oxygen content. This is because the formation of a core or otherwise will affect the mantle oxidation state \citep[e.g.,][]{PP7-licht}. The third scenario can be distinguished from the first two by measurement of \ce{Al2O3} alongside iron silicates. This is because, as an incompatible element, Al will be enhanced if fractional melting is responsible but not if the melt is iron-rich for a different reason. A measurement of corundum in the dust tail of K2-22b may be possible, as discussed in \citet{CamposEstrada23}.

However, truly linking the composition of dust to the planet may still require more work on the outflows. Even if our conclusions hold, the composition of the observed dust may be further altered by physics in the outflow. For instance, refractory species, such as corundum, may condense and rain out. A potentially important effect, as mentioned in \citet{Booth_disint22}, is grain composition changing as different materials condense in subsequent layers. 

Nevertheless, our conclusion, that the material evaporated from the catastrophically evaporating planets matches that melted from the mantle into the lava pool, strengthens the argument for using their dust tails to probe the interior composition of rocky planets.

\subsection{Consequences for atmospheric observations}\label{sec: atm obs}
As mentioned back in \autoref{sec:intro}, it may be possible to detect emission from silicates in lava atmospheres. Our models in \autoref{ch:Mpool sec: likelihood of evolved} imply that most atmospheres should be evolved; thus, they potentially show different features from what one would expect for unevolved atmospheres. \citet{zilinskas22} compared modelled emission spectra of evolved atmospheres to less evolved atmospheres. They found that the features are generally less strong in the evolved case. This is firstly because the abundances of \ce{SiO} and \ce{SiO2} are lower but also because the removal of \ce{Na} decreases the strength of the temperature inversion in the atmosphere. This is because \ce{Na} has high shortwave opacity, which means it absorbs incoming stellar irradiation stronger than infrared radiation from the planet contributing to an atmospheric thermal inversion. There is also clearly an even lower chance of observing emission from \ce{Na} in the evolved case, which was already unlikely. 

As can be seen in \autoref{fig: total_pressures}, evolved atmospheres are predicted to have significantly lower atmospheric pressures than their unevolved counterparts. This may contribute to an explanation of Spitzer non-detections of atmospheres on LHS 3844b \citep{Kreid19LHS} and GJ 1252b \citep{Crossfield_2022}. It will also be significant for the interpretation of future Webb spectra of potential lava atmosphere-bearing planets \citep[e.g.,][]{k2141b_jwst_prop}.

One final consideration for observability is that if traditionally volatile elements (H, C, N, O) are retained in the atmospheres, as discussed in \autoref{sec: volatiles}, then they may make the subtleties of the silicate features even more difficult to discern \citep{piette2023,Zilinskas2023}. 

\section{Summary}\label{ch:Mpool sec:summary}
In this work, we have presented a fairly simple model of the chemical evolution of a lava pool and its atmosphere as the atmosphere is lost, which still captures the essential physics and phenomena. The key new development was the inclusion of melting into the lava pool. The main result is that if sufficient material is removed from the atmosphere, the pool--atmosphere system reaches a steady state where the composition of the atmosphere matches that of  material melted from the mantle (while being very different from the composition of the pool itself). This confirms predictions by \citet{Kite16} but in a framework that is able to explicitly model the process of reaching this steady state due to the inclusion of melting. For reasonable pool compositions, a mass equivalent to $\sim10$--100 pools must be lost to arrive at this state. Significantly, if the mass of the pool is small relative to that of the planet, as it is believed it should be, this state can be reached without the total mass of the planet being altered greatly. It is likely that catastrophically evaporating planets will fall into this regime, meaning the compositions of their tails may directly reflect that of material melted from the mantle, unaltered by different volatilities. Furthermore, since day-to-nightside winds may be able to transport far more material than is lost completely from planets, it may be that even fairly massive planets ($\gtrsim 1 M_\oplus$) have very significant pool evolution. The lower pressures of the evolved atmospheres may make apparent non-detections more likely. Further work is required to understand the consequences of moving so much material to the nightside of a planet.

\section{Data availability}
The code and simulated data underlying this article will be shared on reasonable request to the corresponding author. Any observational or experimental data used are available at the references in the main text.

\section{Acknowledgements}
 A.C. acknowledges the support of an STFC PhD studentship. J.E.O. is supported by the Royal Society through University Research Fellowships. J.E.O. and A.C. have also received funding from the European Research Council (ERC) under the European Union’s Horizon 2020 research and innovation programme (Grant agreement No. 853022, PEVAP). For the purpose of open access, the authors have applied a Creative Commons Attribution (CC-BY) licence to any Author Accepted Manuscript version arising. This work makes use of the Python packages \texttt{Scipy} \citep{scipy2020}, \texttt{Numpy} \citep{numpy}, \texttt{Pandas} \citep{PANDAS_paper,PANDAS_1.5.2} and \texttt{Matplotlib} \citep{matplotlib}.



\bibliographystyle{mnras}
\bibliography{refs}

\begin{thebibliography}{}
\makeatletter
\relax
\def\mn@urlcharsother{\let\do\@makeother \do\$\do\&\do\#\do\^\do\_\do\%\do\~}
\def\mn@doi{\begingroup\mn@urlcharsother \@ifnextchar [ {\mn@doi@} {\mn@doi@[]}}
\def\mn@doi@[#1]#2{\def\@tempa{#1}\ifx\@tempa\@empty \href {http://dx.doi.org/#2} {doi:#2}\else \href {http://dx.doi.org/#2} {#1}\fi \endgroup}
\def\mn@eprint#1#2{\mn@eprint@#1:#2::\@nil}
\def\mn@eprint@arXiv#1{\href {http://arxiv.org/abs/#1} {{\tt arXiv:#1}}}
\def\mn@eprint@dblp#1{\href {http://dblp.uni-trier.de/rec/bibtex/#1.xml} {dblp:#1}}
\def\mn@eprint@#1:#2:#3:#4\@nil{\def\@tempa {#1}\def\@tempb {#2}\def\@tempc {#3}\ifx \@tempc \@empty \let \@tempc \@tempb \let \@tempb \@tempa \fi \ifx \@tempb \@empty \def\@tempb {arXiv}\fi \@ifundefined {mn@eprint@\@tempb}{\@tempb:\@tempc}{\expandafter \expandafter \csname mn@eprint@\@tempb\endcsname \expandafter{\@tempc}}}

\bibitem[\protect\citeauthoryear{{Asimow} \& {Ghiorso}}{{Asimow} \& {Ghiorso}}{1998}]{MELTS2}
{Asimow} P.~D.,  {Ghiorso} M.~S.,  1998, \mn@doi [American Mineralogist] {10.2138/am-1998-9-1022}, \href {https://ui.adsabs.harvard.edu/abs/1998AmMin..83.1127A} {83, 1127}

\bibitem[\protect\citeauthoryear{Bird, Stewart  \& Lightfoot}{Bird et~al.}{2001}]{bird2001transport}
Bird R.,  Stewart W.,   Lightfoot E.,  2001, Transport Phenomena.
Wiley International edition, Wiley, \url {https://books.google.co.uk/books?id=94m7QgAACAAJ}

\bibitem[\protect\citeauthoryear{{Bonsor} \& {Xu}}{{Bonsor} \& {Xu}}{2017}]{Amyreview}
{Bonsor} A.,  {Xu} S.,  2017, in {Pessah} M.,  {Gressel} O.,  eds, Astrophysics and Space Science Library, Vol.~445, Astrophysics and Space Science Library.
p.~229, \mn@doi{10.1007/978-3-319-60609-5_8}

\bibitem[\protect\citeauthoryear{{Booth}, {Owen}  \& {Schulik}}{{Booth} et~al.}{2023}]{Booth_disint22}
{Booth} R.~A.,  {Owen} J.~E.,   {Schulik} M.,  2023, \mn@doi [\mnras] {10.1093/mnras/stac3121}, \href {https://ui.adsabs.harvard.edu/abs/2023MNRAS.518.1761B} {518, 1761}

\bibitem[\protect\citeauthoryear{{Boukar{\'e}}, {Cowan}  \& {Badro}}{{Boukar{\'e}} et~al.}{2022}]{boukare2022}
{Boukar{\'e}} C.-{\'E}.,  {Cowan} N.~B.,   {Badro} J.,  2022, \mn@doi [\apj] {10.3847/1538-4357/ac8792}, \href {https://ui.adsabs.harvard.edu/abs/2022ApJ...936..148B} {936, 148}

\bibitem[\protect\citeauthoryear{{Bourrier} et~al.,}{{Bourrier} et~al.}{2018}]{Bourrier2018}
{Bourrier} V.,  et~al., 2018, \mn@doi [\aap] {10.1051/0004-6361/201833154}, \href {https://ui.adsabs.harvard.edu/abs/2018A&A...619A...1B} {619, A1}

\bibitem[\protect\citeauthoryear{{Bromley} \& {Chiang}}{{Bromley} \& {Chiang}}{2023}]{Bromely-Chiang23}
{Bromley} J.,  {Chiang} E.,  2023, \mn@doi [arXiv e-prints] {10.48550/arXiv.2302.04898}, \href {https://ui.adsabs.harvard.edu/abs/2023arXiv230204898B} {p. arXiv:2302.04898}

\bibitem[\protect\citeauthoryear{{Brouwers}, {Bonsor}  \& {Malamud}}{{Brouwers} et~al.}{2023}]{Brouwers2023}
{Brouwers} M.~G.,  {Bonsor} A.,   {Malamud} U.,  2023, \mn@doi [\mnras] {10.1093/mnras/stac3316}, \href {https://ui.adsabs.harvard.edu/abs/2023MNRAS.519.2646B} {519, 2646}

\bibitem[\protect\citeauthoryear{Buchan, Bonsor, Shorttle, Wade, Harrison, Noack  \& Koester}{Buchan et~al.}{2021}]{buchan2021}
Buchan A.~M.,  Bonsor A.,  Shorttle O.,  Wade J.,  Harrison J.,  Noack L.,   Koester D.,  2021, \mn@doi [Monthly Notices of the Royal Astronomical Society] {10.1093/mnras/stab3624}, 510, 3512

\bibitem[\protect\citeauthoryear{{Campos Estrada}, {Owen}, {Jankovic}, {Wilson}  \& {Helling}}{{Campos Estrada} et~al.}{2024}]{CamposEstrada23}
{Campos Estrada} B.,  {Owen} J.~E.,  {Jankovic} M.~R.,  {Wilson} A.,   {Helling} C.,  2024, \mn@doi [\mnras] {10.1093/mnras/stae095}, \href {https://ui.adsabs.harvard.edu/abs/2024MNRAS.528.1249C} {528, 1249}

\bibitem[\protect\citeauthoryear{{Castan} \& {Menou}}{{Castan} \& {Menou}}{2011}]{castan-menou11}
{Castan} T.,  {Menou} K.,  2011, \mn@doi [\apjl] {10.1088/2041-8205/743/2/L36}, \href {https://ui.adsabs.harvard.edu/abs/2011ApJ...743L..36C} {743, L36}

\bibitem[\protect\citeauthoryear{Chapman \& Cowling}{Chapman \& Cowling}{1990}]{chapman1990}
Chapman S.,  Cowling T.,  1990, The Mathematical Theory of Non-uniform Gases: An Account of the Kinetic Theory of Viscosity, Thermal Conduction and Diffusion in Gases.
Cambridge Mathematical Library, Cambridge University Press, \url {https://books.google.co.uk/books?id=y2Yyy798WzIC}

\bibitem[\protect\citeauthoryear{Chase}{Chase}{1998}]{NIST-JANAF}
Chase M.~W.,  1998, NIST-JANAF thermochemical tables.
American Chemical Society, Washington, DC

\bibitem[\protect\citeauthoryear{Crossfield et~al.,}{Crossfield et~al.}{2022}]{Crossfield_2022}
Crossfield I. J.~M.,  et~al., 2022, \mn@doi [The Astrophysical Journal Letters] {10.3847/2041-8213/ac886b}, 937, L17

\bibitem[\protect\citeauthoryear{{Curry}, {Booth}, {Owen}  \& {Mohanty}}{{Curry} et~al.}{2024}]{CURRY2024}
{Curry} A.,  {Booth} R.,  {Owen} J.~E.,   {Mohanty} S.,  2024, \mn@doi [\mnras] {10.1093/mnras/stae191}, \href {https://ui.adsabs.harvard.edu/abs/2024MNRAS.tmp..190C} {}

\bibitem[\protect\citeauthoryear{{Dang} et~al.,}{{Dang} et~al.}{2021}]{k2141b_jwst_prop}
{Dang} L.,  et~al., 2021, {A Hell of a Phase Curve: Mapping the Surface and Atmosphere of a Lava Planet K2-141b}, JWST Proposal. Cycle 1, ID. \#2347

\bibitem[\protect\citeauthoryear{{Elkins-Tanton} \& {Seager}}{{Elkins-Tanton} \& {Seager}}{2008}]{Elkins-Tanton2008}
{Elkins-Tanton} L.~T.,  {Seager} S.,  2008, \mn@doi [\apj] {10.1086/592316}, \href {https://ui.adsabs.harvard.edu/abs/2008ApJ...688..628E} {688, 628}

\bibitem[\protect\citeauthoryear{{Elkins-Tanton}, {Parmentier}  \& {Hess}}{{Elkins-Tanton} et~al.}{2003}]{Elkins-Tanton2003}
{Elkins-Tanton} L.~T.,  {Parmentier} E.~M.,   {Hess} P.~C.,  2003, \mn@doi [Meteoritics \& Planetary Science] {10.1111/j.1945-5100.2003.tb00013.x}, \href {https://ui.adsabs.harvard.edu/abs/2003M&PS...38.1753E} {38, 1753}

\bibitem[\protect\citeauthoryear{{Fegley} \& {Cameron}}{{Fegley} \& {Cameron}}{1987}]{fegley1987}
{Fegley} B.,  {Cameron} A.~G.~W.,  1987, \mn@doi [Earth and Planetary Science Letters] {10.1016/0012-821X(87)90196-8}, \href {https://ui.adsabs.harvard.edu/abs/1987E&PSL..82..207F} {82, 207}

\bibitem[\protect\citeauthoryear{{Fortney}, {Marley}  \& {Barnes}}{{Fortney} et~al.}{2007}]{Fortney07}
{Fortney} J.~J.,  {Marley} M.~S.,   {Barnes} J.~W.,  2007, \mn@doi [\apj] {10.1086/512120}, \href {https://ui.adsabs.harvard.edu/abs/2007ApJ...659.1661F} {659, 1661}

\bibitem[\protect\citeauthoryear{Frost}{Frost}{1991}]{Frost1991}
Frost B.~R.,  1991, in Lindsley D.~H.,  ed., , Petrologic and Magnetic Significance.
De Gruyter, Berlin, Boston, pp 1--10, \mn@doi{doi:10.1515/9781501508684-004}, \url {https://doi.org/10.1515/9781501508684-004}

\bibitem[\protect\citeauthoryear{Frost \& McCammon}{Frost \& McCammon}{2008}]{annurev-mantle-fugacity}
Frost D.~J.,  McCammon C.~A.,  2008, \mn@doi [Annual Review of Earth and Planetary Sciences] {10.1146/annurev.earth.36.031207.124322}, 36, 389

\bibitem[\protect\citeauthoryear{Ghiorso \& Kress}{Ghiorso \& Kress}{2004}]{Ghiorso+Kress2004}
Ghiorso M.~S.,  Kress V.~C.,  2004, American Journal of Science, 304, 679

\bibitem[\protect\citeauthoryear{{Ghiorso} \& {Sack}}{{Ghiorso} \& {Sack}}{1995}]{MELTS1}
{Ghiorso} M.~S.,  {Sack} R.~O.,  1995, \mn@doi [Contributions to Mineralogy and Petrology] {10.1007/BF00307281}, \href {https://ui.adsabs.harvard.edu/abs/1995CoMP..119..197G} {119, 197}

\bibitem[\protect\citeauthoryear{Guimond, Shorttle, Jordan  \& Rudge}{Guimond et~al.}{2023}]{Guimond_2023}
Guimond C.~M.,  Shorttle O.,  Jordan S.,   Rudge J.~F.,  2023, \mn@doi [Monthly Notices of the Royal Astronomical Society] {10.1093/mnras/stad2486}, 525, 3703

\bibitem[\protect\citeauthoryear{{Harris} et~al.,}{{Harris} et~al.}{2020}]{numpy}
{Harris} C.~R.,  et~al., 2020, \mn@doi [\nat] {10.1038/s41586-020-2649-2}, \href {https://ui.adsabs.harvard.edu/abs/2020Natur.585..357H} {585, 357}

\bibitem[\protect\citeauthoryear{Hindmarsh}{Hindmarsh}{1983}]{hindmarsh83}
Hindmarsh A.~C.,  1983, in Stepleman R.~S.,  ed., Scientific Computing. North-Holland, Amsterdam, pp 55--64

\bibitem[\protect\citeauthoryear{{Hu}, {Ehlmann}  \& {Seager}}{{Hu} et~al.}{2012}]{Hu2012}
{Hu} R.,  {Ehlmann} B.~L.,   {Seager} S.,  2012, \mn@doi [\apj] {10.1088/0004-637X/752/1/7}, \href {https://ui.adsabs.harvard.edu/abs/2012ApJ...752....7H} {752, 7}

\bibitem[\protect\citeauthoryear{Hu et~al.,}{Hu et~al.}{2024}]{55cnce-nat24}
Hu R.,  et~al., 2024, \mn@doi [Nature] {10.1038/s41586-024-07432-x}

\bibitem[\protect\citeauthoryear{{Hunten}, {Pepin}  \& {Walker}}{{Hunten} et~al.}{1987}]{Hunten1987}
{Hunten} D.~M.,  {Pepin} R.~O.,   {Walker} J.~C.~G.,  1987, \mn@doi [\icarus] {10.1016/0019-1035(87)90022-4}, \href {https://ui.adsabs.harvard.edu/abs/1987Icar...69..532H} {69, 532}

\bibitem[\protect\citeauthoryear{Hunter}{Hunter}{2007}]{matplotlib}
Hunter J.~D.,  2007, \mn@doi [Computing in Science \& Engineering] {10.1109/MCSE.2007.55}, 9, 90

\bibitem[\protect\citeauthoryear{{Ito}, {Ikoma}, {Kawahara}, {Nagahara}, {Kawashima}  \& {Nakamoto}}{{Ito} et~al.}{2015}]{Ito2015}
{Ito} Y.,  {Ikoma} M.,  {Kawahara} H.,  {Nagahara} H.,  {Kawashima} Y.,   {Nakamoto} T.,  2015, \mn@doi [\apj] {10.1088/0004-637X/801/2/144}, \href {https://ui.adsabs.harvard.edu/abs/2015ApJ...801..144I} {801, 144}

\bibitem[\protect\citeauthoryear{{Kang}, {Ding}, {Wordsworth}  \& {Seager}}{{Kang} et~al.}{2021}]{Kang2021}
{Kang} W.,  {Ding} F.,  {Wordsworth} R.,   {Seager} S.,  2021, \mn@doi [\apj] {10.3847/1538-4357/abcaa7}, \href {https://ui.adsabs.harvard.edu/abs/2021ApJ...906...67K} {906, 67}

\bibitem[\protect\citeauthoryear{{Kang}, {Nimmo}  \& {Ding}}{{Kang} et~al.}{2023}]{kang2023-TPW}
{Kang} W.,  {Nimmo} F.,   {Ding} F.,  2023, \mn@doi [\apjl] {10.3847/2041-8213/acd691}, \href {https://ui.adsabs.harvard.edu/abs/2023ApJ...949L..20K} {949, L20}

\bibitem[\protect\citeauthoryear{{Kasting}, {Eggler}  \& {Raeburn}}{{Kasting} et~al.}{1993}]{Kasting1993}
{Kasting} J.~F.,  {Eggler} D.~H.,   {Raeburn} S.~P.,  1993, \mn@doi [Journal of Geology] {10.1086/648219}, \href {https://ui.adsabs.harvard.edu/abs/1993JG....101..245K} {101, 245}

\bibitem[\protect\citeauthoryear{Kelemen \& Behn}{Kelemen \& Behn}{2016}]{kelemen2016}
Kelemen P.~B.,  Behn M.~D.,  2016, \mn@doi [Nature Geoscience] {10.1038/ngeo2662}, 9, 197

\bibitem[\protect\citeauthoryear{{Kite}, {Fegley}, {Schaefer}  \& {Gaidos}}{{Kite} et~al.}{2016}]{Kite16}
{Kite} E.~S.,  {Fegley} Bruce J.,  {Schaefer} L.,   {Gaidos} E.,  2016, \mn@doi [The Astrophysical Journal] {10.3847/0004-637X/828/2/80}, \href {https://ui.adsabs.harvard.edu/abs/2016ApJ...828...80K} {828, 80}

\bibitem[\protect\citeauthoryear{Klein}{Klein}{2005}]{klein2005crust}
Klein E.,  2005, The Crust, Vol. 3: Treatise on Geochemistry

\bibitem[\protect\citeauthoryear{Kreidberg et~al.,}{Kreidberg et~al.}{2019}]{Kreid19LHS}
Kreidberg L.,  et~al., 2019, \mn@doi [Nature] {10.1038/s41586-019-1497-4}, 573, 87

\bibitem[\protect\citeauthoryear{{Lammer}, {Kasting}, {Chassefi{\`e}re}, {Johnson}, {Kulikov}  \& {Tian}}{{Lammer} et~al.}{2008}]{Lammer2008}
{Lammer} H.,  {Kasting} J.~F.,  {Chassefi{\`e}re} E.,  {Johnson} R.~E.,  {Kulikov} Y.~N.,   {Tian} F.,  2008, \mn@doi [\ssr] {10.1007/s11214-008-9413-5}, \href {https://ui.adsabs.harvard.edu/abs/2008SSRv..139..399L} {139, 399}

\bibitem[\protect\citeauthoryear{{L{\'e}ger} et~al.,}{{L{\'e}ger} et~al.}{2009}]{leger2009}
{L{\'e}ger} A.,  et~al., 2009, \mn@doi [\aap] {10.1051/0004-6361/200911933}, \href {https://ui.adsabs.harvard.edu/abs/2009A&A...506..287L} {506, 287}

\bibitem[\protect\citeauthoryear{{L{\'e}ger} et~al.,}{{L{\'e}ger} et~al.}{2011}]{leger2011}
{L{\'e}ger} A.,  et~al., 2011, \mn@doi [\icarus] {10.1016/j.icarus.2011.02.004}, \href {https://ui.adsabs.harvard.edu/abs/2011Icar..213....1L} {213, 1}

\bibitem[\protect\citeauthoryear{{Lichtenberg}, {Schaefer}, {Nakajima}  \& {Fischer}}{{Lichtenberg} et~al.}{2023}]{PP7-licht}
{Lichtenberg} T.,  {Schaefer} L.~K.,  {Nakajima} M.,   {Fischer} R.~A.,  2023, in {Inutsuka} S.,  {Aikawa} Y.,  {Muto} T.,  {Tomida} K.,   {Tamura} M.,  eds,  Astronomical Society of the Pacific Conference Series Vol. 534, Protostars and Planets VII. p.~907 (\mn@eprint {arXiv} {2203.10023}), \mn@doi{10.48550/arXiv.2203.10023}

\bibitem[\protect\citeauthoryear{Luger \& Barnes}{Luger \& Barnes}{2015}]{lugerBarnes2015}
Luger R.,  Barnes R.,  2015, \mn@doi [Astrobiology] {10.1089/ast.2014.1231}, 15, 119

\bibitem[\protect\citeauthoryear{McSween, Richardson  \& Uhle}{McSween et~al.}{2003}]{Geo:paths+processes}
McSween H.~Y.,  Richardson S.~M.,   Uhle M.~E.,  2003, Geochemistry: Pathways and Processes, 2 edn.
Columbia University Press, \url {http://www.jstor.org/stable/10.7312/mcsw12440}

\bibitem[\protect\citeauthoryear{{Miguel}, {Kaltenegger}, {Fegley}  \& {Schaefer}}{{Miguel} et~al.}{2011}]{Miguel2011}
{Miguel} Y.,  {Kaltenegger} L.,  {Fegley} B.,   {Schaefer} L.,  2011, \mn@doi [\apjl] {10.1088/2041-8205/742/2/L19}, \href {https://ui.adsabs.harvard.edu/abs/2011ApJ...742L..19M} {742, L19}

\bibitem[\protect\citeauthoryear{Neufeld, Janzen  \& Aziz}{Neufeld et~al.}{2003}]{neufeld72}
Neufeld P.~D.,  Janzen A.~R.,   Aziz R.~A.,  2003, \mn@doi [The Journal of Chemical Physics] {10.1063/1.1678363}, 57, 1100

\bibitem[\protect\citeauthoryear{Nguyen, Cowan, Banerjee  \& Moores}{Nguyen et~al.}{2020}]{nguyen2020}
Nguyen T.~G.,  Cowan N.~B.,  Banerjee A.,   Moores J.~E.,  2020, \mn@doi [Monthly Notices of the Royal Astronomical Society] {10.1093/mnras/staa2487}, 499, 4605

\bibitem[\protect\citeauthoryear{{Palme} \& {O'Neill}}{{Palme} \& {O'Neill}}{2003}]{palme_oneill2003}
{Palme} H.,  {O'Neill} H. S.~C.,  2003, \mn@doi [Treatise on Geochemistry] {10.1016/B0-08-043751-6/02177-0}, \href {https://ui.adsabs.harvard.edu/abs/2003TrGeo...2....1P} {2, 568}

\bibitem[\protect\citeauthoryear{{Perez-Becker} \& {Chiang}}{{Perez-Becker} \& {Chiang}}{2013}]{Perez-Becker13}
{Perez-Becker} D.,  {Chiang} E.,  2013, \mn@doi [\mnras] {10.1093/mnras/stt895}, \href {https://ui.adsabs.harvard.edu/abs/2013MNRAS.433.2294P} {433, 2294}

\bibitem[\protect\citeauthoryear{{Piette}, {Gao}, {Brugman}, {Shahar}, {Lichtenberg}, {Miozzi}  \& {Driscoll}}{{Piette} et~al.}{2023}]{piette2023}
{Piette} A. A.~A.,  {Gao} P.,  {Brugman} K.,  {Shahar} A.,  {Lichtenberg} T.,  {Miozzi} F.,   {Driscoll} P.,  2023, \mn@doi [\apj] {10.3847/1538-4357/acdef2}, \href {https://ui.adsabs.harvard.edu/abs/2023ApJ...954...29P} {954, 29}

\bibitem[\protect\citeauthoryear{{Rappaport} et~al.,}{{Rappaport} et~al.}{2012}]{KIC1255-discov}
{Rappaport} S.,  et~al., 2012, \mn@doi [\apj] {10.1088/0004-637X/752/1/1}, \href {https://ui.adsabs.harvard.edu/abs/2012ApJ...752....1R} {752, 1}

\bibitem[\protect\citeauthoryear{{Rappaport}, {Barclay}, {DeVore}, {Rowe}, {Sanchis-Ojeda}  \& {Still}}{{Rappaport} et~al.}{2014}]{KOI2700b-discov}
{Rappaport} S.,  {Barclay} T.,  {DeVore} J.,  {Rowe} J.,  {Sanchis-Ojeda} R.,   {Still} M.,  2014, \mn@doi [\apj] {10.1088/0004-637X/784/1/40}, \href {https://ui.adsabs.harvard.edu/abs/2014ApJ...784...40R} {784, 40}

\bibitem[\protect\citeauthoryear{{Sanchis-Ojeda} et~al.,}{{Sanchis-Ojeda} et~al.}{2015}]{K2-22b-discov}
{Sanchis-Ojeda} R.,  et~al., 2015, \mn@doi [\apj] {10.1088/0004-637X/812/2/112}, \href {https://ui.adsabs.harvard.edu/abs/2015ApJ...812..112S} {812, 112}

\bibitem[\protect\citeauthoryear{{Schaefer} \& {Fegley}}{{Schaefer} \& {Fegley}}{2009}]{Schaefer_fegley2009}
{Schaefer} L.,  {Fegley} B.,  2009, \mn@doi [\apjl] {10.1088/0004-637X/703/2/L113}, \href {https://ui.adsabs.harvard.edu/abs/2009ApJ...703L.113S} {703, L113}

\bibitem[\protect\citeauthoryear{{Schulik} \& {Booth}}{{Schulik} \& {Booth}}{2023}]{AIOLOS2023}
{Schulik} M.,  {Booth} R.~A.,  2023, \mn@doi [\mnras] {10.1093/mnras/stad1251}, \href {https://ui.adsabs.harvard.edu/abs/2023MNRAS.523..286S} {523, 286}

\bibitem[\protect\citeauthoryear{Solomatov}{Solomatov}{2007}]{SOLOMATOV-chapter}
Solomatov V.,  2007, in Schubert G.,  ed., , Treatise on Geophysics.
Elsevier, Amsterdam, pp 91--119, \mn@doi{https://doi.org/10.1016/B978-044452748-6.00141-3}, \url {https://www.sciencedirect.com/science/article/pii/B9780444527486001413}

\bibitem[\protect\citeauthoryear{{Veras}, {Mustill}  \& {Bonsor}}{{Veras} et~al.}{2024}]{Veras-delivery2024}
{Veras} D.,  {Mustill} A.~J.,   {Bonsor} A.,  2024, \mn@doi [arXiv e-prints] {10.48550/arXiv.2401.08767}, \href {https://ui.adsabs.harvard.edu/abs/2024arXiv240108767V} {p. arXiv:2401.08767}

\bibitem[\protect\citeauthoryear{Virtanen et~al.,}{Virtanen et~al.}{2020}]{scipy2020}
Virtanen P.,  et~al., 2020, \mn@doi [Nature Methods] {10.1038/s41592-019-0686-2}, \href {https://rdcu.be/b08Wh} {17, 261}

\bibitem[\protect\citeauthoryear{{W}es {M}c{K}inney}{{W}es {M}c{K}inney}{2010}]{PANDAS_paper}
{W}es {M}c{K}inney 2010, in {S}t\'efan van~der {W}alt {J}arrod {M}illman eds, {P}roceedings of the 9th {P}ython in {S}cience {C}onference. pp 56 -- 61, \mn@doi{10.25080/Majora-92bf1922-00a}

\bibitem[\protect\citeauthoryear{{Wolf}, {J{\"a}ggi}, {Sossi}  \& {Bower}}{{Wolf} et~al.}{2023}]{VapoRock}
{Wolf} A.~S.,  {J{\"a}ggi} N.,  {Sossi} P.~A.,   {Bower} D.~J.,  2023, \mn@doi [\apj] {10.3847/1538-4357/acbcc7}, \href {https://ui.adsabs.harvard.edu/abs/2023ApJ...947...64W} {947, 64}

\bibitem[\protect\citeauthoryear{{Wright} et~al.,}{{Wright} et~al.}{2023}]{k2-22b_jwst}
{Wright} J.~T.,  et~al., 2023, {Measuring the Interior Composition of a Terrestrial Planet}, JWST Proposal. Cycle 2, ID. \#3315

\bibitem[\protect\citeauthoryear{{Yung}, {Wen}, {Pinto}, {Allen}, {Pierce}  \& {Paulson}}{{Yung} et~al.}{1988}]{Yung1988}
{Yung} Y.~L.,  {Wen} J.-S.,  {Pinto} J.~P.,  {Allen} M.,  {Pierce} K.~K.,   {Paulson} S.,  1988, \mn@doi [\icarus] {10.1016/0019-1035(88)90147-9}, \href {https://ui.adsabs.harvard.edu/abs/1988Icar...76..146Y} {76, 146}

\bibitem[\protect\citeauthoryear{Zahnle \& Kasting}{Zahnle \& Kasting}{1986}]{ZAHNLE1986}
Zahnle K.~J.,  Kasting J.~F.,  1986, \mn@doi [Icarus] {https://doi.org/10.1016/0019-1035(86)90051-5}, 68, 462

\bibitem[\protect\citeauthoryear{{Zahnle}, {Kasting}  \& {Pollack}}{{Zahnle} et~al.}{1990}]{Zahnle1990}
{Zahnle} K.,  {Kasting} J.~F.,   {Pollack} J.~B.,  1990, \mn@doi [\icarus] {10.1016/0019-1035(90)90050-J}, \href {https://ui.adsabs.harvard.edu/abs/1990Icar...84..502Z} {84, 502}

\bibitem[\protect\citeauthoryear{Zhen \& Davies}{Zhen \& Davies}{1983}]{LJparameters}
Zhen S.,  Davies G.~J.,  1983, \mn@doi [physica status solidi (a)] {https://doi.org/10.1002/pssa.2210780226}, 78, 595

\bibitem[\protect\citeauthoryear{{Zilinskas}, {van Buchem, C. P. A.}, {Miguel, Y.}, {Louca, A.}, {Lupu, R.}, {Zieba, S.}  \& {van Westrenen, W.}}{{Zilinskas} et~al.}{2022}]{zilinskas22}
{Zilinskas} M.,  {van Buchem, C. P. A.} {Miguel, Y.} {Louca, A.} {Lupu, R.} {Zieba, S.}  {van Westrenen, W.} 2022, \mn@doi [A&A] {10.1051/0004-6361/202142984}, 661, A126

\bibitem[\protect\citeauthoryear{{Zilinskas}, {Miguel}, {van Buchem}  \& {Snellen}}{{Zilinskas} et~al.}{2023}]{Zilinskas2023}
{Zilinskas} M.,  {Miguel} Y.,  {van Buchem} C.~P.~A.,   {Snellen} I.~A.~G.,  2023, \mn@doi [\aap] {10.1051/0004-6361/202245521}, \href {https://ui.adsabs.harvard.edu/abs/2023A&A...671A.138Z} {671, A138}

\bibitem[\protect\citeauthoryear{{Zuckerman}, {Melis}, {Klein}, {Koester}  \& {Jura}}{{Zuckerman} et~al.}{2010}]{Zuckerman2010}
{Zuckerman} B.,  {Melis} C.,  {Klein} B.,  {Koester} D.,   {Jura} M.,  2010, \mn@doi [\apj] {10.1088/0004-637X/722/1/725}, \href {https://ui.adsabs.harvard.edu/abs/2010ApJ...722..725Z} {722, 725}

\bibitem[\protect\citeauthoryear{pandas~development team}{pandas~development team}{2022}]{PANDAS_1.5.2}
pandas~development team T.,  2022, pandas-dev/pandas: Pandas, \mn@doi{10.5281/zenodo.7344967}, \url {https://doi.org/10.5281/zenodo.7344967}

\bibitem[\protect\citeauthoryear{{van Buchem}, {Miguel}, {Zilinskas}  \& {van Westrenen}}{{van Buchem} et~al.}{2023}]{LavAtmos}
{van Buchem} C. P.~A.,  {Miguel} Y.,  {Zilinskas} M.,   {van Westrenen} W.,  2023, \mn@doi [Meteoritics \& Planetary Science] {10.1111/maps.13994}, \href {https://ui.adsabs.harvard.edu/abs/2023M&PS...58.1149V} {58, 1149}

\bibitem[\protect\citeauthoryear{{van Lieshout} \& {Rappaport}}{{van Lieshout} \& {Rappaport}}{2018}]{disint18}
{van Lieshout} R.,  {Rappaport} S.~A.,  2018, in {Deeg} H.~J.,  {Belmonte} J.~A.,  eds, , Handbook of Exoplanets.
Springer, Cham, p.~1527, \mn@doi{10.1007/978-3-319-55333-7_15}

\bibitem[\protect\citeauthoryear{{van Lieshout}, {Min}  \& {Dominik}}{{van Lieshout} et~al.}{2014}]{vanLieshout14}
{van Lieshout} R.,  {Min} M.,   {Dominik} C.,  2014, \mn@doi [\aap] {10.1051/0004-6361/201424876}, \href {https://ui.adsabs.harvard.edu/abs/2014A&A...572A..76V} {572, A76}

\bibitem[\protect\citeauthoryear{{van Lieshout} et~al.,}{{van Lieshout} et~al.}{2016}]{vanLieshout16}
{van Lieshout} R.,  et~al., 2016, \mn@doi [\aap] {10.1051/0004-6361/201629250}, \href {https://ui.adsabs.harvard.edu/abs/2016A&A...596A..32V} {596, A32}

\bibitem[\protect\citeauthoryear{{van Summeren}, {Conrad}  \& {Gaidos}}{{van Summeren} et~al.}{2011}]{vanSummeren2011}
{van Summeren} J.,  {Conrad} C.~P.,   {Gaidos} E.,  2011, \mn@doi [\apjl] {10.1088/2041-8205/736/1/L15}, \href {https://ui.adsabs.harvard.edu/abs/2011ApJ...736L..15V} {736, L15}

\makeatother
\end{thebibliography}

%%%%%%%%%%%%%%%%%%%%%%%%%%%%%%%%%%%%%%%%%%%%%%%%%%

\appendix
\section{Calculation of theoretical binary diffusion coefficients}\label{app: binary diff}
As seen in \autoref{ch: Mpool sec: pool_results}, the species that make up the atmospheres of interest to this work are not usually gaseous on Earth, and so there is little experimental data available for them. However, one can instead attempt to derive theoretical coefficients from the kinetic theory of gases.

We derived coefficients using Chapman-Enskog theory \citep[e.g.,][]{chapman1990}, which allows the derivation of transport properties as a function of the inter-particle potential between gas particles. The formula for the diffusivity is \citep{bird2001transport}
\begin{equation}
    \mathcal{D}_{ij} = \frac{3}{16} \frac{\sqrt{\frac{2}{\pi} \left(k_BT\right)^3 N_A \left(\frac{1}{m_i}+\frac{1}{m_j}\right)}} {P N_A \sigma_{ij}^2 \Omega_{\mathcal{D},ij}} \; . \label{ch:Mpool , eq: diffusivity}
\end{equation}
Here $N_A$ is Avogadro's number, $T$ and $P$ are the temperature and pressure, $m_{i,j}$ are the molecular masses of the gas species, $\sigma_{ij}$ is the collision cross-section and $\Omega_{\mathcal{D},ij}$ is the `collision integral for diffusion' which depends on the inter-molecular potential. The units for  $\mathcal{D}_{ij}$ are the usual for diffusion coefficients of (length)$^2$ / time.
The diffusivity is related to the binary diffusion coefficient through
\begin{equation}
    b_{ij} = n \mathcal{D}_{ij} \; ,
\end{equation}
where $n$ is the overall number density.

Chapman-Eskog theory strictly only applies to monatomic gases, although it has been shown to produce reasonable results for polyatomic gases too \citep[][\S17.3]{bird2001transport}. As we already have to stretch well beyond experimental data, we restrict calculations to relevant monatomic gases and note that polyatomic gases, if of similar mass, will likely be more readily coupled to escape as they should have larger cross-sections and potentially stronger inter-molecular interactions due to dipoles.

The `collision integral', $\Omega_{\mathcal{D},ij}$ is generally a numerical function, which depends on temperature. If particle interactions are modelled using the Lennard-Jones potential,
\begin{equation}
    \varphi_{ij}(r) = 4\varepsilon_{ij}\left[ \left(\frac{\sigma_{ij}}{r}\right)^{12} - \left(\frac{\sigma_{ij}}{r}\right)^6 \right] \; ,\label{ch:Mpool , eq: LJ}
\end{equation}
the collision integral is a function of $k_BT/\varepsilon_{ij}$. \citet{neufeld72} give an empirical formula for this integral.

The parameters for the interactions between two types of particles can be estimated from the parameters for same-particle interactions using the relations
\begin{equation}
    \sigma_{ij} = \frac{1}{2} \left( \sigma_{i} + \sigma_{j} \right)
\end{equation}
and 
\begin{equation}
    \varepsilon_{ij} = \sqrt{\varepsilon_{i}\varepsilon_{j}} \; .
\end{equation}
Parameters for some relevant atoms (see \autoref{ch: Mpool sec: pool_results}) are shown in \autoref{ch: Mpool tab: gas diff params}.

The form of the equations results in the escape factors having a very weak temperature dependence, so we only plot one temperature in \autoref{fig: escape factors}. This is because the combination of the $T^{3/2} / \Omega_{\mathcal{D}_{ij}}(T)$ in the diffusivity $\mathcal{D}_{ij}$ (\autoref{ch:Mpool , eq: diffusivity}) happens to be $\appropto \, T^2$. Thus $b_{ij} = P/(k_BT) \mathcal{D}_{ij} \appropto \,T$, and so roughly cancels out the $1/T$ factor in \autoref{ch:Mpool eq: two-species escape factor}.

\begin{table}
\resizebox{\linewidth}{!}{%}
\begin{tabular}{|l|c|c|c|c|}\hline
\multirow{2}{*}{\textbf{Parameter [unit]}} & \multirow{2}{*}{\textbf{Definition}} & \multicolumn{3}{c|}{\textbf{Atom}}                  \\ \cline{3-5}
                                   & & \textbf{O} & \textbf{Mg} & \textbf{Fe} \\ \hline
$m_i$ [g mol$^{-1}$] & Atomic mass  & 15.999 & 24.305   & 55.845    \\\hline
$\sigma_i$ [\r{A}] & Lennard-Jones cross section & 3.4281.04 & 2.9234 & 2.3193 \\\hline
\multirow{2}{*}{$\varepsilon_i/k_B$ [K]} & \multirow{2}{*}{$\frac{\text{Lennard-Jones binding energy}}{\text{Boltzman's constant}}$}   & \multirow{2}{*}{118.36}  & \multirow{2}{*}{2071.4} & \multirow{2}{*}{26026.7}  \\ 
& & & & \\\hline
\end{tabular}
}
\caption[Parameters for some relevant atoms]{Atomic parameters for some relevant atoms in lava pool atmospheres. Experimental fits to the parameters of the Lennard-Jones potential (\autoref{ch:Mpool , eq: LJ}) are taken from \citet{LJparameters}. }\label{ch: Mpool tab: gas diff params}
\end{table}

\section{Numerical limit to perfectly reaching equilibrium}\label{ch:Mpool sec:limit to eqm}
\begin{figure}
    \centering
    \includegraphics[width=\linewidth]{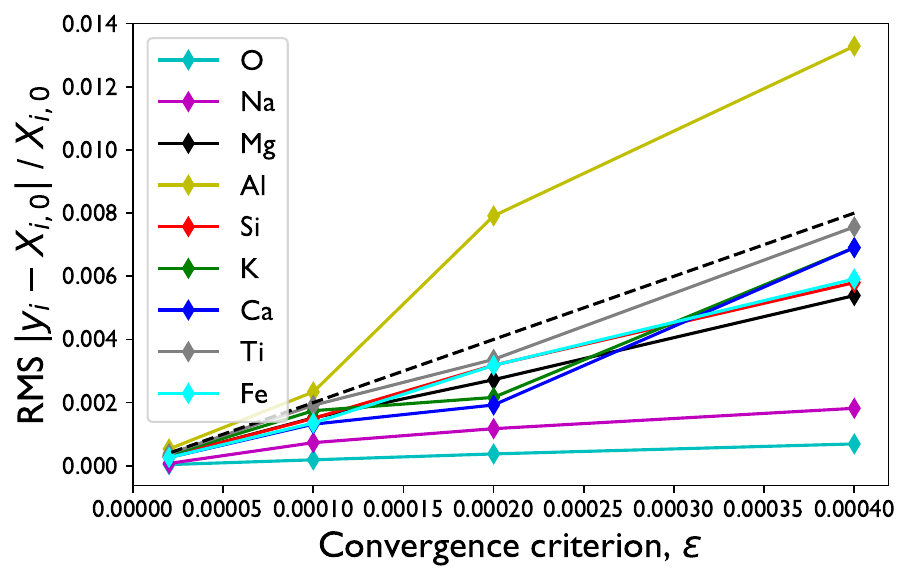}
    \caption[Difference between atmosphere and mantle composition]{The R.M.S. fractional difference between the mantle ($X_{i,m}$) and atmospheric ($Y_i$) elemental abundances for the last 100 steps of 2600 K {\it BSE} pool evolutions run to 500 pools worth of mass removed. The black dashed line shows the theoretical limit to the accuracy of the equilibrium reached, \autoref{ch:Mpool eq:eqm_limit}.}
    \label{fig: convergence_criterion}
\end{figure}

While the compositions are clearly changing very little at late points in the evolution (note also the logarithmic $x$-axis), suggesting an equilibrium, inspection of the lower panel of \autoref{fig: BSE convergence} shows that the final atmosphere and mantle compositions do not perfectly match to arbitrary tolerance. This is because of our stepping algorithm described in \autoref{ch:Mpool sec:step_control}. Since the procedure will take a longer step if it believes equilibrium is achieved up to a given tolerance, the accuracy of the final equilibrium will be limited by this tolerance. Here, we estimate how good the equilibrium can be. 

When the atmospheric concentration is close to the final equilibrium, its composition should be given by
\begin{equation}
    Y_i = X_{i,m} + \delta Y_i \; ,\label{ch:Mpool eq:deviation from eqm}
\end{equation}
where $\delta Y_i$ is some small deviation. Elements are assumed to be in a pseudo-steady state if \autoref{ch:Mpool eq:pseudo_eqm_condition} is satisfied. From Equations \ref{ch:Mpool eq: next step Xi}, \ref{ch:Mpool eq:scaling_vap_to_removed} and \ref{ch:Mpool eq:deviation from eqm}, \autoref{ch:Mpool eq:pseudo_eqm_condition} becomes
\begin{equation}
    \frac{|b X_{i,m} - \xi X_i|}{X_
    i} < \varepsilon \; , \label{ch:Mpool eq:pseudo_eqm_condition2}
\end{equation}
where, as a reminder, $\xi$ and $\varepsilon$ are both non-physical model parameters, the first of which limits the amount of an element that can be removed from the melt per step and the latter of which defines a condition of pseudo-steady state (both are small). Since, in the final state, all the elements will be in a steady state, the earlier subscript $v$ is redundant. We have also dropped the step number labelling.

If all the elements are close to equilibrium, then:
\begin{equation}
    b = \xi\frac{X_i}{Y_i} \left( 1 + \frac{\sum_j \delta Y_j m_j}{\sum_j X_{j,0}m_j} \right) \; ,
\end{equation}
which simply uses Equations \ref{ch:Mpool eq:b_same_pool_mass} and \ref{ch:Mpool eq:deviation from eqm}. Subbing this into \autoref{ch:Mpool eq:pseudo_eqm_condition2} and using \autoref{ch:Mpool eq:deviation from eqm} gives
\begin{equation}
    \left\lvert -\frac{\delta Y_i}{Y_i} + \left(1 - \frac{\delta Y_i}{Y_i}\right) \frac{\sum_j \delta Y_j m_j}{\sum_j X_{j,0}m_j} \right\lvert < \frac{\varepsilon}{\xi} \; .
\end{equation}
$\sum_j \delta Y_j m_j / \sum_j X_{j,0}m_j$ should be $O(\delta Y_i/Y_i)$ and so, ignoring terms of $O(\delta Y_i^2/Y_i^2)$,
\begin{equation}
    \left\lvert \frac{\delta Y_i}{X_{i,m}} \right\lvert \lesssim  \frac{\varepsilon}{\xi} \; . \label{ch:Mpool eq:eqm_limit}
\end{equation}
Thus, one should not expect the final equilibrium to be much better than the right-hand side of this condition.

$| \delta Y_i / X_{i,m}|$ is shown for our fiducial value ($\varepsilon = \num{2e-4}$) in \autoref{fig: BSE convergence} with a dot-dashed line. Meanwhile, \autoref{fig: convergence_criterion} plots $| \delta Y_i / X_{i,m}|$ for all elements at a late stage of the evolution of an initially {\it BSE} pool, for different values of $\varepsilon$, and compares it to $\varepsilon / \xi$ (black dashed line). $\xi$ is fixed at 0.05 for all the models. One can see that the accuracy of the equilibrium does indeed increase as $\varepsilon$, and does so roughly linearly, as also predicted. All of the elements are within the approximate bound for equilibrium except for aluminium, which is the most refractory element, so it should be the last to reach equilibrium. It is, however, within a factor of 2 for all values of $\varepsilon$ (and even closer for the smallest $\varepsilon$.) Thus, it seems reasonable to infer that the composition has reached an equilibrium to as good an accuracy as it might reasonably do.

%%%%%%%%%%%%%%%%%%%%%%%%%%%%%%%%%%%%%%%%%%%%%%%%%%

% Don't change these lines
\bsp	% typesetting comment
\label{lastpage}
\end{document}